\documentclass[lettersize,journal,10pt]{IEEEtran}

\usepackage[margin=1in]{geometry}

\usepackage{graphics} % for pdf, bitmapped graphics files
\usepackage{epsfig} % for postscript graphics files
\usepackage{mathptmx} %  It supersedes both the original times and the mathptm packages.
\usepackage{amsthm}
\usepackage{amsmath} % assumes amsmath package installed
\usepackage{amssymb}  % assumes amsmath package installed
\usepackage{cuted}
\usepackage{float}
\usepackage{color}
\usepackage{url}
\usepackage{bm}

\usepackage[ruled,vlined,linesnumbered,noresetcount]{algorithm2e}
\usepackage{tabularx}

\usepackage{caption,subcaption,cite}

\newtheorem{definition}{Definition}

\makeatletter
\def\ps@IEEEtitlepagestyle{%
  \def\@oddfoot{\mycopyrightnotice}%
  \def\@evenfoot{}%
}
\def\mycopyrightnotice{%
  \begin{minipage}{\textwidth}
  \centering \scriptsize
  Copyright~\copyright~2022 IEEE. Personal use of this material is permitted. Permission from IEEE must be obtained for all other uses, in any current or future media, including reprinting/republishing this material for advertising or promotional purposes, creating new collective works, for resale or redistribution to servers or lists, or reuse of any copyrighted component of this work in other works. 
  \end{minipage}
}
\makeatother

\begin{document}

\title{ADVERT: An Adaptive and Data-Driven Attention Enhancement Mechanism for Phishing Prevention}

\author{Linan Huang, %~\IEEEmembership{Student Member,~IEEE,} 
 Shumeng Jia, 
         Emily Balcetis, and Quanyan Zhu%,~\IEEEmembership{Member,~IEEE}% <-this % stops a space
% <-this % stops a space
\thanks{\textcolor{black}{
Manuscript received 3 February 2022; revised 23 May 2022 and 24 June 2022; accepted 29 June 2022. This work was supported in part by the National Science Foundation (NSF) under Grant ECCS-1847056, Grant CNS-2027884, Grant CNS-1720230, and Grant BCS-2122060; in part by the Army Research Office (ARO) under Grant W911NF-19-1-0041; and in part by the DOE-NE under Grant 20-19829. The associate editor coordinating the review of this manuscript and approving it for publication was Dr. Andrew Beng Jin Teoh. (Corresponding author: Linan Huang.) 
%This work was supported in part by the National Science Foundation (NSF) under Grants ECCS-1847056, CNS-2027884, CNS-1720230, and BCS-2122060; and in part by Army Research Office (ARO) under Grant W911NF-19-1-0041 and DOE-NE under Grant 20-19829.
}}
\thanks{L. Huang,  S. Jia, and Q. Zhu are with the Department
of Electrical and Computer Engineering, New York University,
Brooklyn, NY, 11201, USA. E-mail:\{lh2328, sj3233, qz494\}@nyu.edu} 
\thanks{Emily Balcetis is with the Department of Psychology, New York University, New York, NY, 10003, USA. E-mail:eb107@nyu.edu}
\thanks{Digital Object Identifier 10.1109/TIFS.2022.3189530}
}



\maketitle

\begin{abstract}
Attacks exploiting the \textit{innate} and the \textit{acquired} vulnerabilities of human users have posed severe threats to cybersecurity. 
This work proposes ADVERT, a \textit{human-technical solution} that generates adaptive visual aids in real-time to prevent users from inadvertence and reduce their susceptibility to phishing attacks. 
Based on the eye-tracking data, we extract \textit{visual states} and \textit{attention states} as system-level sufficient statistics to characterize the user's visual behaviors and attention status. 
By adopting a data-driven approach and two learning feedback of different time scales, this work \textcolor{black}{lays out} a theoretical foundation to \textit{analyze}, \textit{evaluate}, and particularly \textit{modify} \textcolor{black}{humans' attention processes} while \textcolor{black}{they vet and recognize phishing emails.}  
%make security decisions of phishing recognition. 
We corroborate the \textit{effectiveness}, \textit{efficiency}, and \textit{robustness} of ADVERT through a case study based on the data set collected from human subject experiments conducted at New York University. 
The results show that the visual aids can statistically increase the attention level and improve the accuracy of phishing recognition from $74.6\%$ to a minimum of $86\%$. 
The meta-adaptation can further improve the accuracy to $91.5\%$ (resp. $93.7\%$) in less than $3$ (resp. $50$) tuning stages. 
\end{abstract}


\section{Introduction}
\IEEEPARstart{H}{uman} is often considered the weakest link in cybersecurity. 
Adversaries can exploit human errors and vulnerabilities to launch deceptive attacks \textcolor{black}{(e.g., social engineering and phishing)} that lead to information leakages and data breaches. Moreover, these attacks often serve as the initial stages of sophisticated attacks \textcolor{black}{(e.g., supply chain attacks and advanced persistent threats) that} inflict tremendous damage on critical infrastructures. 
%How can these vulnerabilities lead to cybersecurity threats?
%Why  is it hard to counter deception for human vulnerabilities?
We classify human vulnerabilities into \textit{innate vulnerabilities} (e.g., bounded attention and rationality) and \textit{acquired vulnerabilities} (e.g., lack of security awareness and \textcolor{black}{incentives}). 
Previous works have \textcolor{black}{mitigated the} acquired vulnerabilities through security training\cite{aldawood2018educating}, rule enforcement \cite{alotaibi2016information}, and incentive designs \cite{huang2021duplicity,huang2022zetar}, but these methods are less than sufficient to deal with the innate ones, especially due to the unpredictability and heterogeneity of human behaviors.
% through short-term training or rules.  
%(e.g. picking the length of the passwords, checking for mistakes, multi-step authentication, etc. ). 
 %In addition, the unpredictability and heterogeneity of human behaviors make it even harder to mitigate innate vulnerabilities.
%Many security solutions have mainly focused on the acquired vulnerabilities, which are insufficient to mitigate the security risk of deceptive attacks exploiting the innate human vulnerabilities. 
To this end, there is a need for \textit{security-assistive technologies} to deter and adaptively correct the user misbehavior resulting from the innate vulnerabilities. 

\begin{figure*}[h]
\centering
\includegraphics[width=0.65 \textwidth]{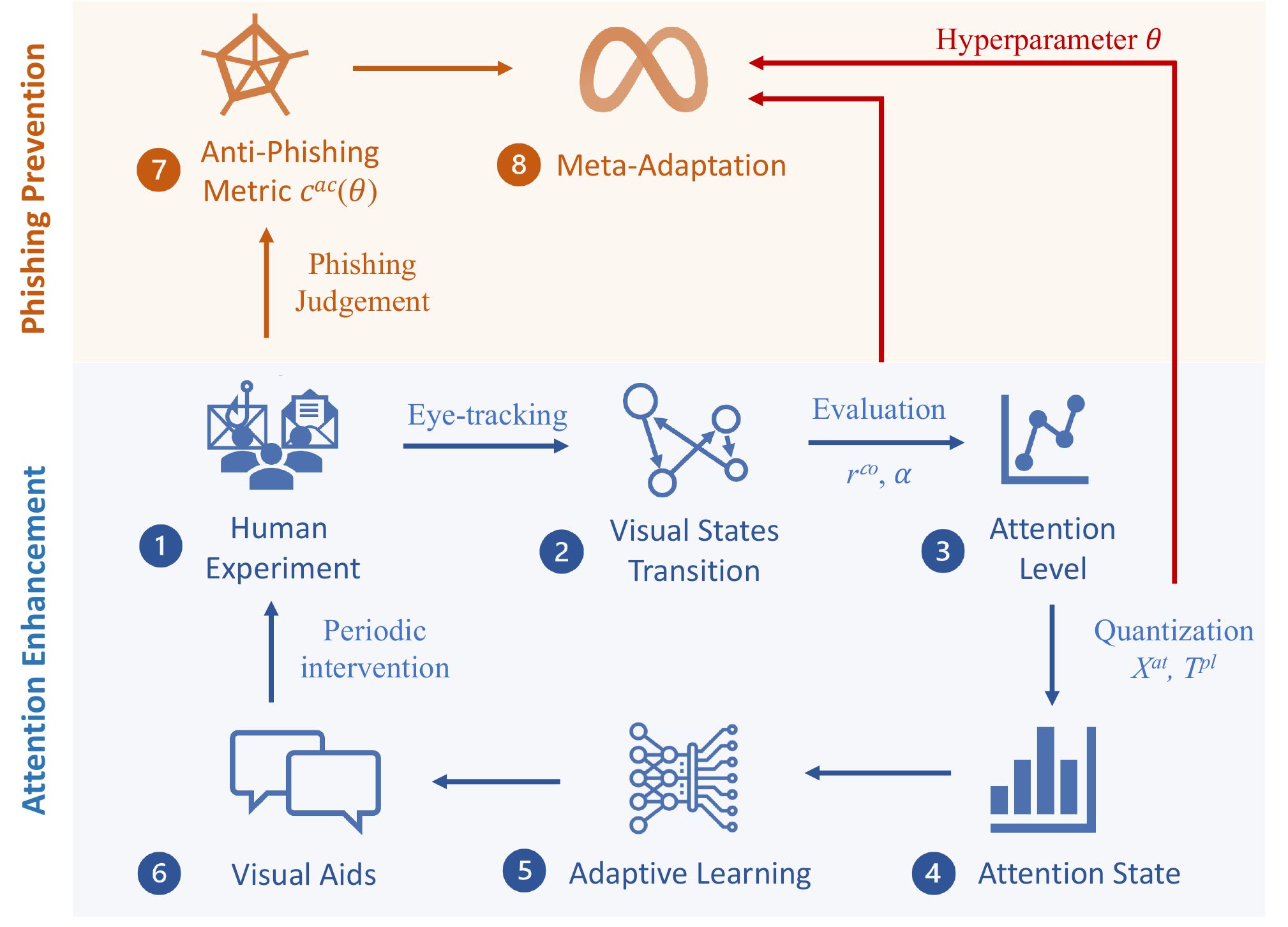} %,height=.5\columnwidth
\caption{ 
The design diagram of ADVERT. 
The adaptive learning loops of the attention enhancement mechanism and the phishing prevention mechanism are highlighted using juxtaposed blue and orange backgrounds, respectively. 
Since a user needs to persistently pay attention to an email to make a phishing judgment, the meta-adaptation feedback in orange updates less frequently than the feedback of attention enhancement in blue. 
}
\label{fig:assistiveSecurityDiag}
\end{figure*}

%The inattention is an innate vulnerability. Only trained professional can control their attention skillfully, e.g. operators, etc.
In this work, we focus on inattention, one type of innate human vulnerability, and use phishing email as a prototypical scenario to explore the users' visual behaviors when they determine whether a \textcolor{black}{sequence of} emails is secure or not. 
Based on the users' eye-tracking data and phishing recognition results, we develop  ADVERT\footnote{ADVERT is an acronym for ADaptive Visual aids for Efficient Real-time security-assistive Technology.} to provide a human-centric data-driven attention enhancement mechanism for phishing prevention. 
In particular, ADVERT enables an adaptive visual-aid generation to guide and sustain the users' attention to the right content of an email and consequently makes users less likely to fall victim to phishing. 
The design of the ADVERT contains two feedback loops of attention enhancement and phishing prevention at short and long time scales, respectively, as shown in Fig. \ref{fig:assistiveSecurityDiag}. 

%, which is commonly exploited by phishers and scammers. 

The bottom part of  Fig. \ref{fig:assistiveSecurityDiag} in blue illustrates the design of adaptive visual aids (e.g., highlighting, warnings, and educational messages) to engage human users in email \textcolor{black}{vetting}. 
First, as a human user reads emails and judges whether they are phishing or legitimate, a covert eye-tracking system can record the user's eye-gaze locations and pupil sizes in real-time. 
Second, based on the eye-tracking data, we abstract the email's Areas of Interest (AoIs), e.g., title, hyperlinks, attachments, etc.,  and develop a Visual State (VS) transition model to characterize the eye-gaze dynamics. 
Third, we develop system-level attention metrics to evaluate the user's attention level based on the VS transition trajectory. 
Then, we quantize the attention level to obtain the Attention State (AS) and develop adaptive learning algorithms to generate visual aids as feedback of the AS. 
The visual aids change the user's hidden cognitive states and lead to the set of eye-tracking data with different patterns of VS transition and AS, which then updates the design of visual aids and enhances attention iteratively. 
%In contrast to low-level eye-tracking metrics related to fixations, saccades, regressions, and backtracks in the previous works  \cite{beymer2005webgazeanalyzer,ramkumar2020eyes}, we propose system-level security metrics to evaluate the magnitude and the frequency of the user's mind-wandering behaviors. 

The attention enhancement loop serves as a stepping-stone to achieving the ultimate goal of phishing prevention. 
The orange background in the top part of  Fig. \ref{fig:assistiveSecurityDiag} illustrates how we tune the hyperparameters in the attention enhancement loop to safeguard users from phishing emails. First, we create a metric to evaluate the user's accuracy in phishing recognition under the current attention enhancement mechanism.  
Then, we iteratively revise the hyperparameters to achieve the highest accuracy. 
Since the accuracy evaluation depends on the implementation of the entire attention enhancement loop, the evaluation is costly and time-consuming. Thus, we leverage Bayesian Optimization (BO) to propose an efficient meta-level tuning algorithm \textcolor{black}{that improves} the accuracy. 
%to improve the accuracy.  %set the optimal attention enhancement loop for

%feature and outcome
The contributions of this work are threefold. 
First, we provide a holistic model of the human-in-the-loop system for email vetting and phishing recognition. 
By abstracting the complex human processes of sensing, thinking, and acting as a stochastic feedback control system of various parameters, we establish a system-level characterization of human attention and security judgment. 
Such characterization focuses on the interaction between the human and the technical systems, especially the inputs (e.g., visual aids) and the outputs (e.g., gaze locations, attention status, and security decisions) of the human system. 
Moreover, we propose new attention metrics to quantify the impact of hidden attention status on observable performance metrics, e.g., \textcolor{black}{accuracy of recognizing phishing}. 
These metrics enable a real-time \textcolor{black}{modification} of the human attention process through the adaptive visual-aid generation. 

Second, we provide an adaptive technology called ADVERT to counteract inattention and improve the human recognition of phishing attacks. 
Two algorithms are developed to illustrate the design, where the \textit{individual adaptation algorithm} improves the visual aid design for each individual user, and the \textit{population adaptation algorithm} further learns the optimal visual aid for the user population.  
%, which will ultimately lead to an open-source software. 
Since the data-driven approach achieves customized solutions in terms of the users and the content of the emails, ADVERT can be applied to various security threat scenarios caused by inattention. 
Since the feedback learning framework enables an adaptive and systematic design of the optimal visual aids, ADVERT can be applied with insufficient domain knowledge. 

Finally, we corroborate the \textit{effectiveness}, \textit{efficiency}, and \textit{robustness} of ADVERT through a case study based on the data set collected from human subject experiments conducted at New York University \cite{cox2020stuck}. 
The results show that the visual aids can sufficiently enhance the attention level and improve the accuracy of phishing recognition from $74.6\%$ to a minimum of $86\%$. When we further tune the hyperparameters, we manage to improve the accuracy of phishing recognition from  $86.8\%$ to  $93.7\%$ in less than $50$ tuning stages, while the largest accuracy improvement happens within $3$ tuning stages. 
The results have also provided insights and guidance for the ADVERT design; e.g., the attention threshold for visual-aid selection (resp. the period length for visual-aid generation) has a small (resp. periodic) impact on phishing recognition. 

% In \cite{dhamija2006phishing}, the authors have identified three reasons why phishing works. They are Lack of Knowledge, visual deception, and bounded attention. 
% The last two require people to pay attention to the right place with sufficient long gaze time to process the information and identify the visual deception. 
% Some deception can be identified with a glance, for example, Images masking underlying text. Once the people look at the hidden link, he knows the deception. 
% Some need more time to process, for example, use 1 for l, people may need to gaze at it for sufficient long time to know the deception. 

\subsection{Notations and Organization of the Paper}
Throughout the paper, we use \textcolor{black}{subscripts} to index time and stages.  
Calligraphic letter $\mathcal{S}$ defines a set and $|\mathcal{S}|$ represents its cardinality. 
The indicator function $\mathbf{1}_{\{A\}}$ takes value $1$ if condition $A$ is true and value $0$ if $A$ is false. 
The rest of the paper is organized as follows. 
The related works are presented in Section \ref{sec:related works}. 
We elaborate on the two feedback loops of Fig. \ref{fig:assistiveSecurityDiag} in Section \ref{sec:BenchmarkModel} and \ref{sec:mata-learning}, respectively. 
Section \ref{sec:case study} presents a case study of ADVERT for email vetting and phishing recognition. \textcolor{black}{Section \ref{sec:limitations} discusses the limitations, and} Section \ref{sec:conclusion} concludes the paper.

\section{Related works}
\label{sec:related works}
\subsection{Phishing Attack Detection and Prevention}
Phishing is the act of masquerading as a legitimate entity to serve malware or steal credentials. 
The authors in \cite{dhamija2006phishing} have identified three human vulnerabilities that make humans the unwitting victims of phishing. 
\begin{itemize}
    \item Lack of knowledge for computer system security; e.g., \url{www.ebay-members-security.com} does not belong to \url{www.ebay.com}. 
    \item Inadequacy to identify visual deception; e.g., the phishing email can contain an image of a legitimate hyperlink, but the image itself serves as a hyperlink to a malicious site.  %\cite{dhamija2006phishing}. 
    A human cannot identify the deception by merely looking at it. 
    %judge whether a hyperlink is an image by merely looking at it. 
    %the legitimate hyperlink is an image that itself serves as a hyperlink to a malicious site. 
    \item Lack of attention (e.g., careless users fail to notice the phishing indicators\textcolor{black}{, including} spelling errors and grammar mistakes) 
    and \textit{inattentional blindness} (e.g., users focusing on the main content fail to perceive unloaded logos in a phishing email \cite{WinNT1}).  
\end{itemize}

Many works have attempted to mitigate the above three human vulnerabilities to prevent phishing attacks.  
First, security education and anti-phishing training, e.g., role-playing phishing simulation games \cite{wen2019hack} and fake phishing attacks \cite{dodge2007phishing}, have been used to compensate for the user's lack of security knowledge and increase users' security awareness. 
Second, detection techniques based on visual similarities \cite{jain2017phishing} and machine learning \cite{khonji2013phishing} have been applied to help users identify visual deception. 
\textcolor{black}{Modern web browsers and email clients also provide security indicators (e.g., the protocol used, the domain name, and the SSL/TLS certificate) to assist users in decision-making \cite{kelley2016real}.} 
Third, passive warnings (i.e., do not block the content-area) and active warnings (i.e., prohibits the user from viewing the content-data) have been \textcolor{black}{developed} empirically to draw users' attention and prevent them from falling victim to phishing \cite{egelman2008you,khonji2013phishing}. 
Our work lays out a foundation to compensate for the third human vulnerability of inattention systematically and quantitatively. 

\subsection{Counterdeception Technologies}
Adversarial cyber deception has been a long-\allowbreak standing problem.  It is easy for an attacker to deceive yet much more difficult for regular users to identify the deception given the universal human vulnerabilities. 
Previous works have mainly focused on \textit{human solutions} (e.g., security training \cite{aldawood2018educating}) or \textit{technical solutions} (e.g., defensive deception technologies \cite{al2019autonomous,huang2020dynamic,pawlick2021game}), to deter, detect, and respond to deceptive attacks.
%Defensive deception technologies such as honeypots \cite{bringer2012survey} and moving target defense \cite{jajodia2011moving} %neutralize the advantages
This work focuses on designing a \textit{human-technical solution} through eye-tracking data, visual aids, and learning techniques to counteract adversarial cyber deception. 
%human-centric technical solutions 

Biosensors, \textcolor{black}{including} eye trackers and electroencephalogram (EEG) devices, provide a window into an analytical understanding of human perception and cognition to enhance security and privacy \cite{katsini2020role}. 
In particular, researches have investigated the users' gaze behaviors and attention when reading Uniform Resource Locators (URLs) \cite{ramkumar2020eyes}, phishing webs \cite{miyamoto2015eye}, and phishing emails \cite{cox2020stuck,mcalaney2020understanding,xiong2017domain}. 
These works illustrate the users' visual processing of phishing contents \cite{ramkumar2020eyes,miyamoto2015eye,pfeffel2019user,mcalaney2020understanding} and the effects of visual aids \cite{xiong2017domain}. 
The authors in \cite{miyamoto2015eye} further establish correlations between eye movements and phishing identification to estimate the likelihood that users may fall victim to phishing attacks. 
Compared to these works that \textit{analyze} human perception, we use eye-tracking data to \textit{design} visual aids and  \textcolor{black}{\textit{modify}} the human perception process for better security decisions. %Moreover, we define proper metrics to measure the user's attention level and the accuracy of phishing identification, enabling the optimal design of the visual support system to counteract visual deception and phishing.
\textcolor{black}{
Moreover, we use biometric data at different granularities. 
Compared to previous works that exploit the \textit{statistics} of the biometric data (e.g., the number of fixations and gaze duration distributions), we use the \textit{dynamic transitions} of the eye-tracking data to extract attention metrics for corrective measures in \textit{real-time}. 
}

\subsection{Human Vulnerability \textcolor{black}{Quantification and  Learning}} %learning
Human plays significant roles in cybersecurity. It is challenging to model, quantify, and affect human behaviors and their mental processes such as reasoning, perception, and cognition. 
\textcolor{black}{
Therefore, various modeling and learning approaches are developed to mitigate human vulnerabilities in cyberspace, as shown in the following two paragraphs, respectively.}

\textcolor{black}{The authors in \cite{canfield2016quantifying,canfield2018setting} use Signaling Detection Theory (SDT) to quantify phishing susceptibility and prioritize behavioral interventions for reducing phishing risk, respectively. 
Adopting SDT, they treat the phishing risk management as a \textit{vigilance task}, where individuals monitor their environment to distinguish signals (i.e., phishing emails) from noises (i.e., legitimate emails). 
Their approaches investigate phishing on a detailed level based on varying factors, including task, individual, and environmental ones.  
%We do not explicitly study how these factors affect performance. 
We adopt a system-level characterization, where system-scientific tools such as feedback, Reinforcement Learning (RL), and \textcolor{black}{BO} are used to adapt to these varying factors.}  

Due to the modeling challenges and the unpredictability, RL \cite{HUANG2022} has been used to characterize and mitigate human vulnerabilities, including bounded rationality \cite{shi2019exploring}, prospect theory \cite{sanjab2020game}, incompliance \cite{huang2021duplicity}, and bounded attention \cite{RN661,huang2021radams}. 
Using \textcolor{black}{RL} to detect, evaluate, and compensate for risks induced by human vulnerabilities is still in its infancy, but it is a promising direction as RL provides a quantitative and adaptive solution. 
%Other machine learning techniques, including support vector machines (SVM), random forests, and neural networks

\section{Attention Enhancement Mechanism}
\label{sec:BenchmarkModel}
\begin{figure}[h]
\centering
\includegraphics[width=1\columnwidth]{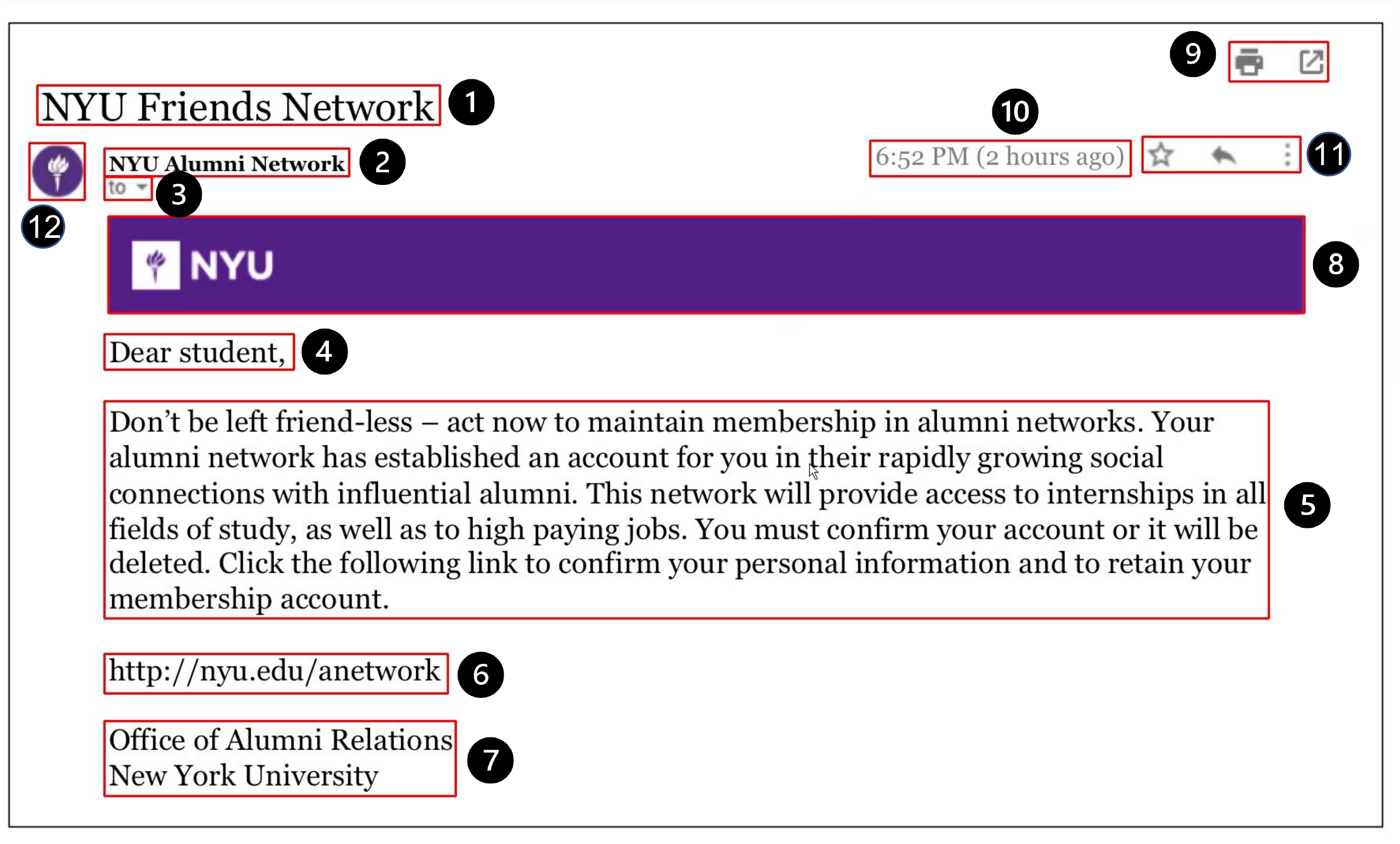}
\caption{
A sample email with $12$ AoIs. 
In sequence, they are the email's title, the sender's information, the receiver's information, the salutation, the \textcolor{black}{main content}, the URL, the sender's signature, the organization logo, the `print' and `share' buttons, the timestamp,  the `bookmark' and `forward' buttons, and the sender's profile picture. 
The AoI partition in red boxes and their index numbers in black circles are invisible to users. 
}
\label{fig:sampleEmail}
\end{figure}

As illustrated by Step $1$ of Fig. \ref{fig:assistiveSecurityDiag}, we consider a group of $M$ human users who vet a list of $N$ emails and classify them as phishing or legitimate. 
As a user $m\in \mathcal{M}:=\{1,\cdots,M\}$ reads an email $n\in \mathcal{N}:=\{1,\cdots,N\}$ on the screen for a duration of $T_m^n$, the eye-tracking device records the vertical and the horizontal coordinates of his eye gaze point in real-time.  
To compress the sensory outcomes and facilitate RL-driven attention enhancement solutions, we aggregate potential gaze locations (i.e., pixels on the screen) into a finite number of $I$ non-overlapping Areas of Interest (AoIs) as shown in Fig. \ref{fig:sampleEmail}. 
We index each potential AoI by $i\in \mathcal{I}:=\{1,2,...,I\}$. 

Each email does not need to contain all the AoIs, and the AoI partition remains unknown to the users. 
Previous works \cite{mcalaney2020understanding,ramkumar2020eyes,miyamoto2015eye} have identified the role of AoIs in helping human users recognize phishing, and different research goals can lead to different AoI partitions.  
%is not unique and should be processed based on the research goal. 
For example, the \textcolor{black}{main content} AoI (i.e., area $5$ in Fig. \ref{fig:sampleEmail}) can be divided into finer AoIs based on the phishing indicators such as misspellings, grammar mistakes, and threatening sentences.  %and monetary information. 
We refer to all other areas in the email (e.g., blank areas)  as the \textit{uninformative area}. 
When the user's eyes  move off the screen during the email vetting process, no coordinates of the gaze location are available. We refer to these off-screen areas as the \textit{distraction area}. 

\subsection{Visual State Transition Model}
\label{sec:VS transition model}
\begin{figure}[h]
\centering
\includegraphics[width=1\columnwidth]{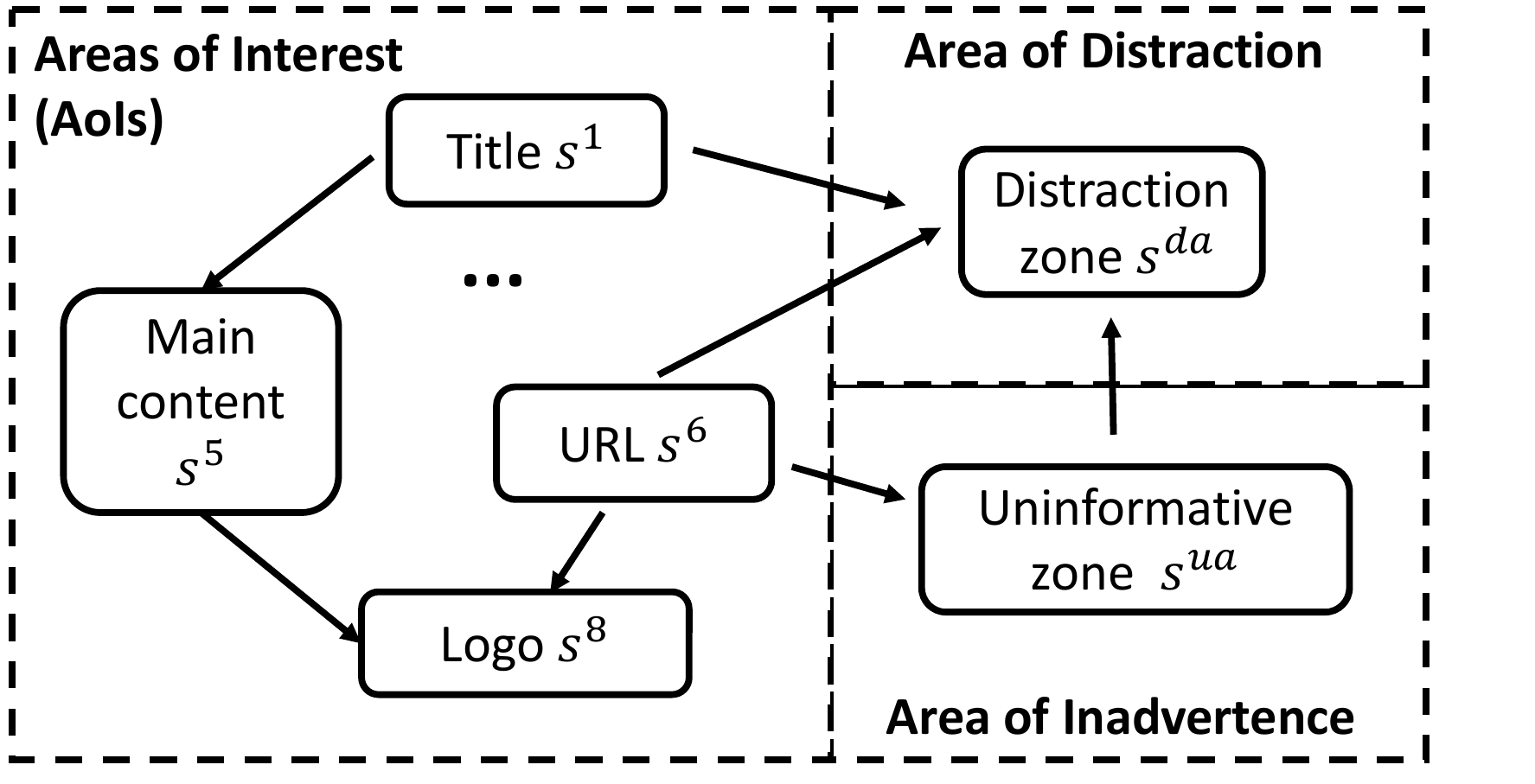} %,height=.5\columnwidth
\caption{ 
Transitions among VSs in $\mathcal{S}$. The VS indices are consistent with \textcolor{black}{the AoI indices in} Fig. \ref{fig:sampleEmail}. 
}
\label{fig:stateTrans}
\end{figure}

As illustrated by Step $2$ \textcolor{black}{in} Fig. \ref{fig:assistiveSecurityDiag}, we establish the following transition model based on the AoI to which the user's gaze location belongs at different times. 
We define $\mathcal{S}:= \{s^i\}_{i\in \mathcal{I}} \cup \{s^{ua},s^{da}\}$ as the set of  $I+2$ \textit{Visual States (VSs)}, where $s^i$ represents the $i$-th AoI; $s^{ua}$ represents the \textit{uninformative area}; and $s^{da}$ represents the \textit{distraction area}. 
We provide an example transition map of these VSs in Fig. \ref{fig:stateTrans}. 
The links represent the potential shifts of the gaze locations during the email reading process; e.g., \textcolor{black}{a user} can shift his focus from the title to the main content or the distraction area. 
We omit most links for illustration purposes; e.g., it is also possible for a user to regain attention to the AoIs from distraction or inadvertence. 

\begin{figure}[h]
%\raggedright
\centering
\includegraphics[width=1\columnwidth]{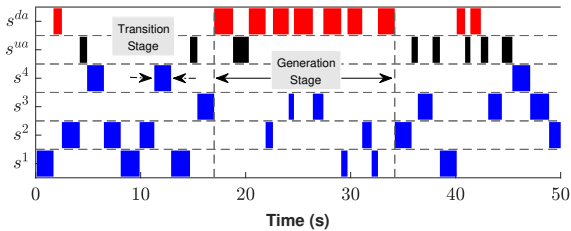} %,height=.5\columnwidth
\caption{ 
An exemplary \textcolor{black}{VS} transition trajectory $[s_t]_{t\in [0,T_m^n]}$. The $x$-axis and the $y$-axis represent $T_m^n=50$ seconds and $I+2=6$ \textcolor{black}{VSs}, respectively. 
We denote \textcolor{black}{VSs}  $s^{da}$, $s^{ua}$, and $\{s^i\}_{i\in \mathcal{I}}$ in red, black, and blue, respectively. 
Each generation stage \textcolor{black}{can} contain different numbers of transition stages. 
}
\label{fig:gazepath}
\end{figure}

We denote $s_t\in \allowbreak \mathcal{S}$ as the VS of user $m\in \mathcal{M}$ vetting email $n\in \mathcal{N}$ at time $t\in \allowbreak [0,T_m^n]$. In this work, we do not distinguish among human users concerning their attention processes while they read different emails. 
Then, each user's gaze path during the interval $[0,T_m^n]$ can be characterized as the same stochastic process $[s_t]_{t\in [0,T_m^n]}$. 
The stochastic transition of the VSs divides the entire time interval $[0,T_m^n]$ into different \textit{transition stages}. 
We visualize an exemplary VS transition trajectory $[s_t]_{t\in [0,T_m^n]}$ in Fig. \ref{fig:gazepath} under $I=4$ AoIs and $T_m^n=50$ seconds. 
As denoted by the colored squares, $40$ VSs arrive in sequence, which results in $40$ discrete transition stages. 
%Compared to the gaze plot in Fig. \ref{fig:attentionphishing} that records the gaze locations across the entire email area, the gaze path in Fig. \ref{fig:gazepath} aggregates these gaze locations into merely $6$ VSs, enabling the real-time attention evaluation and phishing recognition. 

\subsection{Feedback Visual-Aid Design} 
\label{sec:Visual support system}
Propel visual aids can help guide and sustain the users' attention.  
Previous works have proposed different classes of visual aids to enhance phishing recognition, including highlights of contents \cite{xiong2017domain,lin2011does}, warnings of suspicious hyperlinks and attachments \cite{egelman2008you,akhawe2013alice}, and anti-phishing educational messages \cite{sheng2010falls}. 
These potential classes of visual aids construct the visual-aid library denoted as a finite set $\mathcal{A}$. 
 
As illustrated by Step $6$ \textcolor{black}{in} Fig. \ref{fig:assistiveSecurityDiag}, different visual aids can affect the users' visual behaviors. 
The influence, however, can be beneficial (e.g., timely highlights prevent users from mind-wandering) or detrimental (e.g., extensive highlights make humans weary and less attentive to the AoIs). 
\textcolor{black}{
The effectiveness of visual aids for preventing phishing may not be straightforward, especially under different environmental (e.g., security indicator designs) and human factors (e.g., users' security knowledge and prior trust) \cite{kelley2016real}. 
In this paper, we focus on adapting visual aids to the human visual attention.} 
%Due to the unpredictability and heterogeneity of human behaviors and their mental processes, there lacks mature theories or design rules to generate the most beneficial visual aids directly under different conditions. 
%Moreover, the visual aid should adapt to the human visual attention that changes during the email vetting.  
%Therefore, w
We apply \textcolor{black}{RL} to learn the dynamic design of visual aids based on the real-time evaluation of the user's attention status detailed in Section \ref{sec:attentionEva}. 

The sequence of adaptive visual aids is generated with a period of length $T^{pl}$, and we refer to the time interval between every two visual aids as the \textit{generation stage} indexed by $k\in \allowbreak \mathcal{K}_m^n:= \{1,2,  \cdots, \allowbreak K_m^n\}$, where $K_m^n$ is the maximum generation stage during $[0,T_m^n]$; i.e., $K_m^n T^{pl}\leq T_m^n$ and $(K_m^n+1)T^{pl}\geq T_m^n$. 
Then, we denote $a_k\in\mathcal{A}$ as the visual aid at the $k$-th generation stage. 
Fig. \ref{fig:gazepath} illustrates how visual aids affect the transition of \textcolor{black}{VSs} in $K_m^n=3$ generation stages divided by the two vertical dashed lines. 
During the second generation stage, an improper visual aid leads to more frequent transitions to the distraction area and also a longer sojourn time at the \textcolor{black}{VS} $s^{da}$. 
On the contrary, the proper visual aids during the first and the third generation stages engage the users and extend their attention spans, i.e., the amount of time spent on AoIs before a transition to  $s^{da}$ or  $s^{ua}$. 

\subsection{Evaluation of Attention Status}
\label{sec:attentionEva}

From the VS transition trajectory (e.g., Fig. \ref{fig:gazepath}), we aim to construct the \textit{Attention State (AS)} used as the feedback value for the adaptive visual-aid design. 
We define $\mathcal{X}$ as the set of all possible attention states. 
%At each generation stage $k\in \mathcal{K}_m^n$, we obtain the AS $x_k\in \mathcal{X}$ from the VS transition history $[s_t]_{t\in [0,(k-1)T^{pl}]}$ based on the AS generation function $f$. 
Previous works (e.g., \cite{pfeffel2019user,mcalaney2020understanding}) have defined attention metrics based on the AoIs, \textcolor{black}{including} the proportion of time spent on each AOI, gaze duration means, fixation count, and average duration.  
Compared to these \textit{detailed-level} metrics extracted directly from raw eye-gaze data, we propose the following \textit{system-level} metric of attention level based on the VS transition history as \textcolor{black}{will be} shown in Section \ref{sec:average attention level}. 
%AS generation function based on the VS transition history to quantify the attention level defined in Section \ref{sec:average attention level}. 
Such system-level metric serves as sufficient statistics to effectively characterize the attention status. Moreover, it preserves the users' privacy \textcolor{black}{because} the raw data of gaze locations can reveal sensitive information about their biometric identities, including gender, age, and ethnicity \cite{liebling2014privacy, kroger2019does}. 

To this end, we assign scores to each \textcolor{black}{VS} in Section \ref{sec:rewardassigment} to evaluate the user's attention (e.g., gaze at AoIs) and inattention (e.g., gaze at uninformative and distraction areas). 
The scores can be determined manually based on the expert recommendation and empirical studies (e.g., \cite{pfeffel2019user}), or based on other biometric data (e.g., the pupil sizes in Fig. \ref{fig:pupilsize}). 
Moreover, we can apply \textcolor{black}{BO} for further fine-tuning of these scores as shown in Section \ref{sec:hyperparameter Update}. 

\subsubsection{Concentration Scores and Decay Rates}
\label{sec:rewardassigment}

Both the gaze location and the gaze duration matter in the identification of phishing attacks.
For example, at the first glance, users cannot distinguish  the spoofed email address  `\url{paypa1@mail.paypaI.com}' from the authentic one `\url{paypal@mail.paypal.com}' while a guided close look reveals that the lower case letter `\url{l}' is replaced by the number `\url{1}' and the capital letter `\url{I}'. 
Therefore, we assign a \textit{concentration score} $r^{co}(s)\in \mathbb{R}$ to characterize the sustained attention associated with \textcolor{black}{VS} $s\in \mathcal{S}$.  
Since the amount of information that a user can extract from a VS $s\in \mathcal{S}$ is limited, we use  an exponential decay rate of $\alpha(s)\in \mathbb{R}^+$ to penalize the effect of concentration score as time elapses. 
Different \textcolor{black}{VSs} can have different concentration scores and decay rates. 
For example, the \textcolor{black}{main content} AoI \textcolor{black}{(i.e., area $5$ in Fig. \ref{fig:sampleEmail})} usually contains more information than other AoIs, and  an extended attention span extracts more information (e.g., the substitution of letter `\url{l}' into `\url{I}') to identify the phishing email. Thus, the \textcolor{black}{main content} AoI turns to have a high concentration score and a low decay rate, which is corroborated in Table \ref{table:AoIscore} based on the data set collected from human experiments \cite{cox2020stuck} as \textcolor{black}{will be} shown in Section \ref{sec:case study}. 

\subsubsection{Cumulative Attention Level}
\label{sec:average attention level}
\begin{figure}[h]
\centering
\includegraphics[width=1\columnwidth]{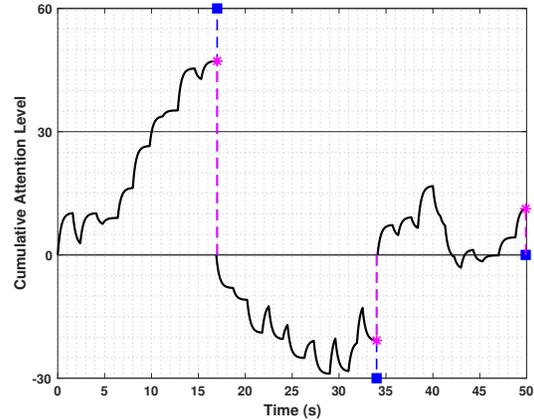} %,height=.5\columnwidth
\caption{ 
The user's cumulative attention level $v_k(t-(k-1)T^{pl}), k\in \mathcal{K}_m^n, t\in [(k-1)T^{pl},kT^{pl}]$, over $K_m^n=3$ generation stages in $T_m^n=50$ seconds. 
The horizontal lines quantize $v_k(t)$ into $X=4$ values that form the finite set $\mathcal{X}=\{-30,0,30,60\}$. The purple star and the blue square denote the values of $\bar{v}_k \cdot T^{pl}$ and $\bar{v}^{qu}_{k} \cdot T^{pl}$, respectively,  at each generation stage $k\in \mathcal{K}_m^n$. 
}
\label{fig:averagereward}
\end{figure}

% Based on the sets of scores associated with $s\in \mathcal{S}$, we define the cumulative reward $ {u}_{k}(s,t)$ at time $t\in [(k-1)T^{pl},kT^{pl}]$ over generation stage $k\in \mathcal{K}_m^n$ as
% \begin{equation*}
%     \begin{split}
%              {u}_k(s,t)  := \int_{(k-1)T^{pl}}^{t} r^{co}(s) 
%     \cdot e^{-\alpha(s) (\tau-(k-1)T^{pl})} \cdot \mathbf{1}_{\{ s_{\tau}=s  \}} d\tau. 
%     \end{split}
% \end{equation*}
We construct the metric for attention level illustrated \textcolor{black}{by} Step $3$ \textcolor{black}{in} Fig. \ref{fig:assistiveSecurityDiag} as follows. 
Let $W_k\in \mathbb{Z}^+$ be the total number of transition stages contained in generation stage  $k\in \mathcal{K}_m^n$. 
Then, we define  $t_k^{w_k}, w_k\in \{1,2,\cdots,W_k\}$, as the duration of the $w_k$-th transition stage in the $k$-th generation stage. 
Take the gaze path in Fig. \ref{fig:gazepath} as an example, the first generation stage contains $w_1=12$ transition stages and the first $7$ transition stages last for a total of $\sum_{w_1=1}^7 t_1^{w_1}=10$ seconds.  
Based on the sets of scores associated with $s\in \mathcal{S}$, we compute the cumulative reward $ {u}^{w_k}_k(s,t)$ at time $t$ of the $w_k$-th transition stage in the $k$-th generation stage as 
$
      {u}^{w_k}_k(s,t)  = \int_{0}^{t} r^{co}(s) e^{-\alpha(s) \tau}  \cdot \mathbf{1}_{\{ s=s^{\tau}  \}}d\tau , 0\leq t\leq t_k^{w_k}. 
$
At generation stage $k$, we define $\bar{w}_k^t$ as the latest transition stage before time $t$, i.e., $\sum_{w_k=1}^{\bar{w}_k^t}t_k^{w_k}\leq t$ and $\sum_{w_k=1}^{\bar{w}_k^t+1}t_k^{w_k}> t$. 
Then, we define the user's \textit{Cumulative Attention Level (CAL)} $v_{k}(t)$ over time interval $[(k-1)T^{pl},t]$ at generation stage $k\in \mathcal{K}_m^n$ as the following cumulative reward 
\begin{equation}
    v_{k}(t):=\sum_{s\in \mathcal{S}} \sum_{w_k=1}^{\bar{w}_k^t}  {u}^{w_k}_k(s,t) , 0 \leq t\leq T^{pl},   %=\sum_{w_k=1}^{W_k} t_k^{w_k}. 
\end{equation}
%Note that this time is the relative time within generation stage $k$. It is not absoluate time. 
We visualize the CAL of $K_m^n=3$ generation stages in Fig. \ref{fig:averagereward} based on the gaze path in Fig. \ref{fig:gazepath}. 
%There are $K=3$ generation stages during which the value of $v_k(t)$ increases and decreases with time $t$ when $s_t\in \{s^i\}_{i\in \mathcal{I}}$ and $s_t\in \{s^{ua},s^{da}\}$, respectively. 

Since $v_k(t)$ is bounded for all $k\in \mathcal{K}_m^n, t\in [0,T^{pl}]$, 
we can quantize it into $X$ finite values to construct the set $\mathcal{X}$ of the attention states illustrated by Step $4$ \textcolor{black}{in} Fig. \ref{fig:assistiveSecurityDiag}. 
We represent the quantized value of $v_k(t)\in \mathbb{R}$ as $v^{qu}_{k}(t)\in \mathcal{X}$ for all $k\in \mathcal{K}_m^n, t\in [0,T^{pl}]$, and define the \textcolor{black}{Average Attention Level (AAL) and Quantized Average Attention Level (QAAL)} 
%average attention level and quantized  average attention level 
for each generation stage in Definition \ref{def:averageAttentionLevel}.  

\begin{definition}%[\textbf{Quantized Average Attention Level}]
\label{def:averageAttentionLevel}
Let $\bar{v}_k\in \mathbb{R}$ and $\bar{v}^{qu}_{k}\in \mathcal{X}$ denote the user's Average Attention Level (AAL) and Quantized Average Attention Level (QAAL) over generation stage $k\in \mathcal{K}_m^n$, respectively.  
They are measured by the improvement \textcolor{black}{in} CAL and the quantized value of the CAL improvement per unit time, i.e., $\bar{v}_k:=v_k(T^{pl})/T^{pl}$ and  $\bar{v}^{qu}_{k}:=v^{qu}_{k}(T^{pl})/T^{pl}$, respectively. 
\end{definition}

\subsection{Q-Learning via Consolidated Data} 
\label{sec:Q-learning}
In Section \ref{sec:Q-learning}, We elaborate on the adaptive learning block \textcolor{black}{(i.e., Step $5$ in Fig. \ref{fig:assistiveSecurityDiag})}. 
%Since the inspection time of each user reading one email is not sufficiently long, we apply Q-learning on the consolidated data of a group of email inspection data. 
Since the inspection time of a user reading one email is not sufficiently long, we consolidate a group of email inspection data to learn the optimal visual-aid generation policy over a population. 

The QAAL $\bar{v}^{qu}_{k}\in \mathcal{X}$ represents the {attention state} at the generation stage $k\in \mathcal{K}_m^n$. 
Since the goal is to enhance the user's attention represented by the CAL, the reward function $R: \mathcal{X}\times\mathcal{A} \mapsto \mathbb{R}$ should be monotone concerning the value of $\bar{v}^{qu}_{k}$, e.g., $R(\bar{v}^{qu}_{k},a_k):=\bar{v}^{qu}_{k}, \forall a_k\in \mathcal{A}$. 
In this work, we assume that each visual aid $a_k\in \mathcal{A}$ exerts the same statistical effect on the attention process regardless of different users and emails. 
Thus, we can consolidate the data set of $\bar{M}\in \{1,\cdots,M\}$ users and $\bar{N}\in \{1,\cdots,N\}$ emails\footnote{When sufficiently large data sets are available, we can carefully choose these $\bar{M}$ users to share similar attributes (e.g., ages, sexes, races, etc.) and these $\bar{N}$ emails to belongs to the same categories (e.g., business or personal emails).} to learn the optimal visual-aid generation policy $\sigma\in \Sigma: \mathcal{X}\mapsto \mathcal{A}$ in a total of $\bar{K}:=\sum_{m=1}^{\bar{M}} \sum_{n=1}^{\bar{N}} K_m^n$ stages. 
With a given discounted factor $\beta\in (0,1)$, the expected long-term objective can be represented as 
$\max_{\sigma\in \Sigma} \mathbb{E} [\sum_{k=1}^{\bar{K}} (\beta)^k \cdot R(\bar{v}^{qu}_{k}, \sigma(\bar{v}^{qu}_{k}))]$. 

 The $Q$-table $[Q_k(\bar{v}^{qu}_k,a_k)]_{\bar{v}^{qu}_k\in \mathcal{X}, a_k\in \mathcal{A}}$ represents the user's attention pattern at generation stage $k\in \mathcal{\bar{K}}:=\{1,\cdots,\bar{K}\}$, i.e., the estimated payoff of applying visual aid $a_k\in \mathcal{A}$ when the attention state is $\bar{v}^{qu}_k\in \mathcal{X}$.  
 Let the sequence of learning rate $\gamma_k(\bar{v}^{qu}_k,a_k)$ satisfy $\sum_{k=0}^{\infty} \gamma_k(\bar{v}^{qu}_k,a_k)=\infty$ and $\sum_{k=0}^{\infty} (\gamma_k(\bar{v}^{qu}_k,a_k))^2<\infty$ for all $\bar{v}^{qu}_k\in \mathcal{X},a_k\in \mathcal{A}$.   
Then, we can update the attention pattern at each generation stage $k\in \mathcal{\bar{K}}$ as follows, i.e., 
\begin{equation}
\label{eq:Qlearning}
    \begin{split}
      & Q_{k+1}(\bar{v}^{qu}_k,\sigma_k(\bar{v}^{qu}_k))=Q_k(\bar{v}^{qu}_k,\sigma_k(\bar{v}^{qu}_k))  
      \\& \quad\quad\quad  + \gamma_k(\bar{v}^{qu}_k,\sigma_k(\bar{v}^{qu}_k)) 
      \cdot [R(\bar{v}^{qu}_{k},\sigma_k(\bar{v}^{qu}_k)) 
      \\& \quad\quad\quad
      +\beta \max_{a\in \mathcal{A}} Q_k(\bar{v}^{qu}_{k+1},a)-Q_k(\bar{v}^{qu}_k,\sigma_k(\bar{v}^{qu}_k))  ],    
    \end{split}
\end{equation}
% \begin{equation}
%     \begin{split}
%     \label{eq:Qlearning}
%       & Q_{k+1}(\bar{v}^{qu}_k,\sigma_k(\bar{v}^{qu}_k))=(1-\gamma_k(\bar{v}^{qu}_k,\sigma_k(\bar{v}^{qu}_k))) Q_k(\bar{v}^{qu}_k,\sigma_k(\bar{v}^{qu}_k)) \\&   + \gamma_k(\bar{v}^{qu}_k,\sigma_k(\bar{v}^{qu}_k)) [R(\bar{v}^{qu}_{k},\sigma_k(\bar{v}^{qu}_k))+\beta \max_{a\in \mathcal{A}} Q_k(\bar{v}^{qu}_{k+1},a) ],    
%     \end{split}
% \end{equation}
where the visual-aid generation policy $\sigma_k(\bar{v}^{qu}_k)$ at \textcolor{black}{generation stage} $k\in \mathcal{\bar{K}}$ is an $\epsilon_k$-greedy policy; i.e., with probability $\epsilon_k\in [0,1]$, the visual aid $a_k$ is selected randomly from $\mathcal{A}$ and with probability $1-\epsilon_k$, the optimal visual aid $a^{*}_{k}\in \textrm{arg}\max_{a\in\mathcal{A}} Q_k(\bar{v}^{qu}_{k},a)$ is implemented. 
To obtain a convergent \textcolor{black}{visual-aid generation policy}, the value of $\epsilon_k$ gradually decreases from $1$ to $0$. 

%Since the visual aids are generated periodically at the end of each generation stage, there is one-step delay of the Q-evaluation %take about this in the algorithm of case study

% The user's privacy is preserved because our learning method uses the visual aids as the feedback of the high-level attention states at generation stages rather than the low-level visual states at transition stages.  %take about this in the previous sections.

\section{Phishing Prevention Mechanism}
\label{sec:mata-learning}
The attention enhancement mechanism in Section \ref{sec:BenchmarkModel} tracks the attention process in real-time to enable the adaptive visual-aid generation. 
By properly modifying the user's attention and engaging him in vetting emails, the attention enhancement mechanism serves as a stepping-stone to achieving the ultimate goal of phishing prevention. 
Empirical evidence and observations have shown that a high attention level, or mental arousal, does not necessarily yield good performance \cite{posner2016attention}. 
\textcolor{black}{In the specific task of phishing recognition, recent works  \cite{nasser2020role,ackerley2022errors}  have also identified curvilinear relationships between phishing recognition accuracy and critical attentional factors, including a participant's cue utilization, cognitive reflection, and cognitive load.} 
Thus, besides attention metrics, e.g., the AAL, we need to design anti-phishing metrics to measure the users' performance of phishing recognition as \textcolor{black}{will be} shown in Section \ref{sec:evaluation of phishing}. 

In Section \ref{sec:hyperparameter Update}, we develop an efficient meta-level algorithm to tune the hyperparameters \textcolor{black}{(e.g., the period length $T^{pl}$ of the visual-aid generation, the number of attention states $X$, the attention scores $r^{co}(s),\alpha(s), \forall s\in \mathcal{S}$, etc.)} in the attention enhancement mechanism. 
We denote these hyperparameters as one $d$-dimensional variable $\theta=[T^{pl},X, [r^{co}(s)]_{s\in \mathcal{S}},[\alpha(s)]_{s\in \mathcal{S}}]\in \mathbb{R}^d$, where $d=2+2|\mathcal{S}|$. 
Let the $i$-th element $\theta^i$ be upper and lower bounded by $\bar{\theta}^i $ and $ \underline{\theta}^i$, respectively. 
Thus, $\theta\in \Theta^d:=\{[\theta^i]_{i\in \{1,\cdots ,d\}}\in \mathbb{R}^d| \underline{\theta}^i \leq \theta^i\leq  \bar{\theta}^i  \}$. 
%Finally, we evaluate the efficiency of the algorithm in Section \ref{sec:evalationofBO}. 

\subsection{Metrics for Phishing Recognition}
\label{sec:evaluation of phishing}
As illustrated by Step $7$ in Fig. \ref{fig:assistiveSecurityDiag}, we provide a metric to evaluate the outcome of the users' phishing identification under a given hyperparameter $\theta\in \Theta^d$. 
After vetting email $n\in \{1,\cdots,\bar{N}\}$, the user $m\in \{1,\cdots,\bar{M}\}$ judges the email to be phishing or legitimate. 
The binary variable $z_m^n(\theta)\in \{z^{co},z^{wr}\}$ represents whether the judgment is correct (denoted by $z^{co}$) or not (denoted by $z^{wr}$). 
We can reshape the two-dimension index $(m,n)$ as a one-dimension index $\hat{n}$ and rewrite $z_m^n(\theta)$ as $z_{\hat{n}}(\theta)$. 
Once these users have judged in total of $N^{bo}$ emails, we define the
%the accuracy of phishing recognition as 
following metric $c^{ac}\in \mathcal{C}:\Theta^d \mapsto [0,1]$ to evaluate the \textit{accuracy} of phishing recognition, i.e., 
\begin{equation}
\label{eq:accuracy metric}
  c^{ac}(\theta):=\frac{1}{N^{bo}}  \sum_{\hat{n}=1}^{N^{bo}} |\mathbf{1}_{\{ z_{\hat{n}}(\theta)=z^{co}\}}|, \forall \theta\in \Theta^d.  %=h(\{y_n,z_n\}_{n=1}^N)
\end{equation}
% \begin{equation*}
% \label{eq:accuracy metric}
%   c^{ac}(\theta):=\frac{1}{N^{bo}}  \sum_{n=1}^{N^{bo}} |\mathbf{1}_{\{ z^{\theta}_n=z^{ph}\}} -\mathbf{1}_{\{ y^{\theta}_n=y^{no}\}}|, \forall \theta\in \Theta^d.  %=h(\{y_n,z_n\}_{n=1}^N)
% \end{equation*}
% We can also choose the following metrics to measure the accuracy. 
% %Based on the $N$ judgments and their true labels, we list some possible accuracy metrics as follows. 
% \begin{itemize}
%     \item True positive rate $c^{tp}(\theta):=$ \\
%     $ \frac{\sum_{n=1}^N \mathbf{1}_{\{ z^{\theta}_n=z^{ph}, y^{\theta}_n=y^{ph}\}} }{ \sum_{n=1}^N \mathbf{1}_{\{ z^{\theta}_n=z^{ph}, y^{\theta}_n=y^{ph}\}} +\sum_{n=1}^N \mathbf{1}_{\{ z^{\theta}_n=z^{no}, y^{\theta}_n=y^{ph}\}}  }$. 
%     \item  Precision $c^{pr}(\theta):=$ \\
%     $ \frac{\sum_{n=1}^N \mathbf{1}_{\{ z^{\theta}_n=z^{ph}, y^{\theta}_n=y^{ph}\}} }{ \sum_{n=1}^N \mathbf{1}_{\{ z^{\theta}_n=z^{ph}, y^{\theta}_n=y^{ph}\}} +\sum_{n=1}^N \mathbf{1}_{\{ z^{\theta}_n=z^{ph}, y^{\theta}_n=y^{no}\}}  }$.
%     \item F-score $c^{fs}(\theta):=2c^{tp}(\theta)/(1+c^{tp}(\theta)/c^{pr}(\theta))$. 
% \end{itemize}

The goal is to find the optimal hyperparameter $\theta^*\in \Theta^d$ to maximize the accuracy of phishing identification, i.e., $\theta^*\in \textrm{arg}\max_{\theta\in \Theta^d} c^{ac}(\theta)$. 
However, we cannot know the value of $c^{ac}(\theta)$ for a $\theta\in \Theta^d$ a priori until we implement this hyperparameter $\theta$ in the attention enhancement mechanism.  
The implemented hyperparameter affects the adaptive visual-aid generation that changes the user's attention and the anti-phishing performance metric $c^{ac}(\theta)$. 
Since the experimental evaluation at a given $\theta\in \Theta^d$ is time-consuming, we present an algorithm in Section \ref{sec:hyperparameter Update} to determine how to choose and update the hyperparameter to maximize the detection accuracy. 

\subsection{Efficient Hyperparameter Tuning}
\label{sec:hyperparameter Update}
\begin{figure}[h]
\centering
\includegraphics[width=1\columnwidth]{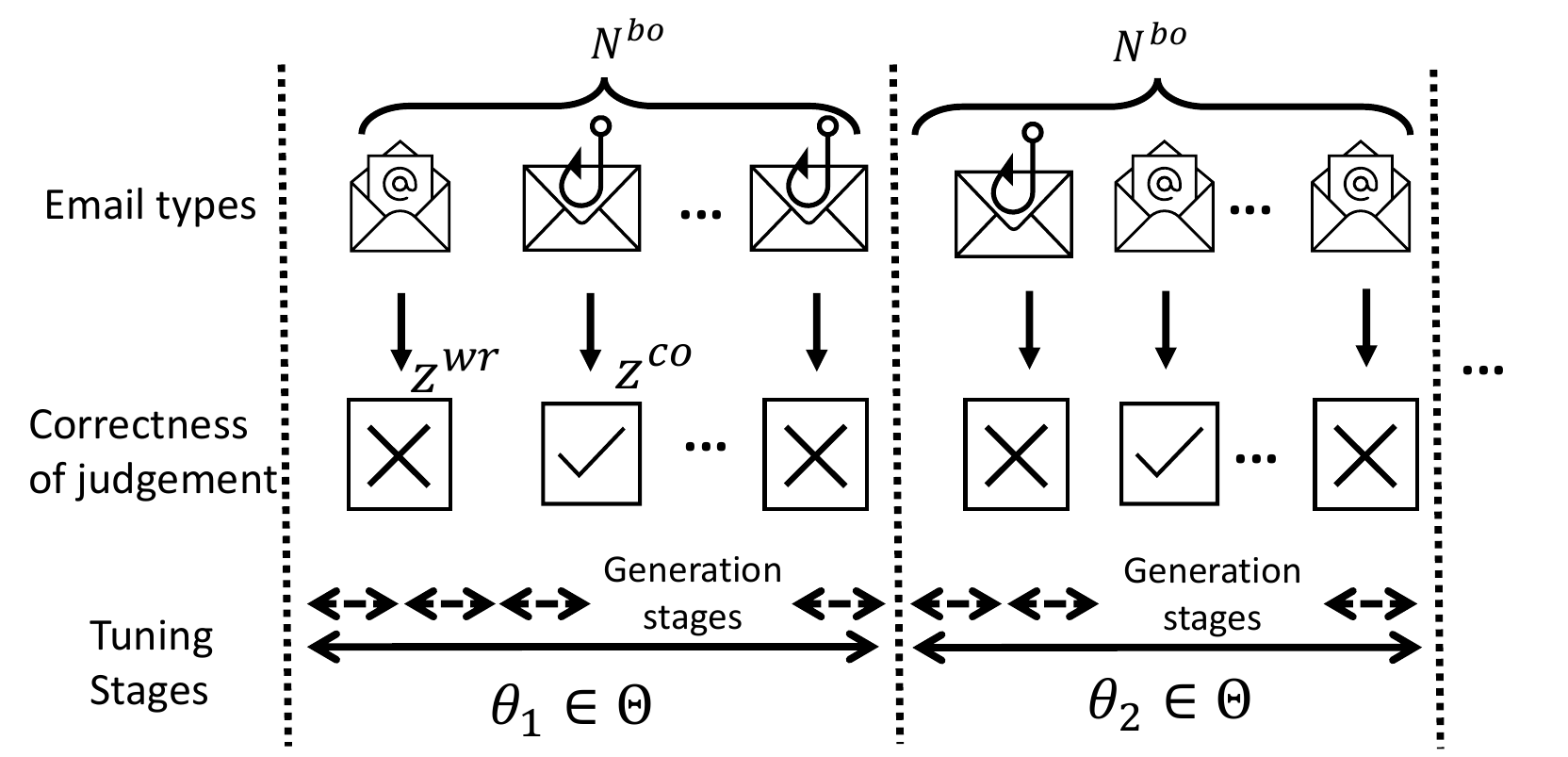} %,height=.5\columnwidth
\caption{
Hyperparameter tuning based on the user's phishing recognition. 
Each tuning stage consists of $N^{bo}$ emails and contains several generation stages. 
}
\label{fig:metalearningDiag}
\end{figure}

We illustrate the meta-adaptation (i.e., Step $8$ in Fig. \ref{fig:assistiveSecurityDiag}) in Section \ref{sec:hyperparameter Update}. 
As illustrated in Fig. \ref{fig:metalearningDiag}, we refer to the duration of every $N^{bo}$ security decisions as a \textit{tuning stage}.  
Consider a time and budget limit that restricts us to conduct $L$ tuning stages in total. We denote $\theta_l$ as the hyperparameter at the $l$-th tuning stage where $l\in \mathcal{L}:=\{1,2,\cdots,L\}$. 
Since each user's email inspection time is different, each tuning stage can contain different numbers of generation stages. 

To find the optimal hyperparameter $\theta^*\in \Theta^d$ within $L$ \textcolor{black}{tuning stages} is challenging. The empirical methods (e.g., a naive grid search \textcolor{black}{or a} random search over $\Theta^d\subset \mathbb{R}^d$) become inefficient when $d>1$. 
\textcolor{black}{BO} \cite{frazier2018bayesian} provides a systematic way to  update the hyperparameter and balance between exploration and exploitation. 
BO consists of a Bayesian statistical model of the objective function $c^{ac}\in \mathcal{C}$ and an acquisition function for deciding the hyperparameter to implement at the next tuning stage. 
%By assuming the unknown objective function $c^{ac}$ is Lipschitz-continuous, i.e., there exists a constant $w^{lc}$ such that $c^{ac}(\theta)-c^{ac}(\bar{\theta})\leq w^{lc}\cdot dx(\theta-\bar{\theta}), \forall \theta,\bar{\theta}\in \Theta^d$. 
The statistical model of $c^{ac}\in \mathcal{C}$ is a Gaussian process $\mathcal{N}(\mu^0,\Sigma^0)$ with a mean function $\mu^0(\theta)=\bar{\mu}^0$ and covariance function or kernel $\Sigma^0(\theta,\bar{\theta})  =  \lambda^0 \cdot \exp(\sum_{i=1}^d \lambda^i (\theta^i-\bar{\theta}^{i})^2)$ for all $\theta,\bar{\theta} \in \Theta^d$, where $\bar{\mu}^0$, $\lambda^0$ and $\lambda^i, i\in \{1,2,\cdots,d\}$, are parameters of the kernel. 
The kernel $\Sigma^0$ is required to be positive semi-definite and has the property that the points closer in the input space are more strongly correlated. 
For any $l\in \mathcal{L}$, we define three shorthand notations $\mu^0(\theta_{1:l}):=[\mu^0(\theta_{1}), \cdots, \mu^0(\theta_{l})]$, $c^{ac}(\theta_{1:l}):=[c^{ac}(\theta_1), \cdots, c^{ac}(\theta_l)]$, and 
%$\Sigma^0(\theta_{1:l},\theta_{1:l})):=[\Sigma^0(\theta_1,\theta_1),\cdots, \Sigma^0(\theta_1,\theta_l);\cdots ;\Sigma^0(\theta_l,\theta_1),\cdots, \Sigma^0(\theta_l,\theta_l)]$
\begin{equation*}
\Sigma^0(\theta_{1:l},\theta_{1:l}):=
\begin{bmatrix}
& \Sigma^0(\theta_1,\theta_1) & \cdots & \Sigma^0(\theta_1,\theta_l)\\
&\vdots  &\ddots  & \vdots\\
&\Sigma^0(\theta_l,\theta_1) &\cdots & \Sigma^0(\theta_l,\theta_l)
\end{bmatrix}.
\end{equation*} 
Then, the evaluation vector of $l\in\mathcal{L}$ elements is assumed to be multivariate Gaussian distributed, i.e.,  
$c^{ac}(\theta_{1:l})\sim \mathcal{N}(\mu^0(\theta_{1:l}),\Sigma^0(\theta_{1:l},\theta_{1:l}))$. 
Conditioned on the values of $\theta_{1:l}$, we can infer the value of $c^{ac}(\theta)$ at any other $\theta\in \Theta\setminus \{\theta_{{l}'}\}_{{l}'\in \{1,\cdots,l\}}$ by Bayesian rule, i.e., 
\begin{equation}
    \begin{split}
    \label{eq:conditionDis}
        c^{ac}(\theta) | c^{ac}(\theta_{1:l}) \sim \mathcal{N}(\mu^n(\theta), (\Sigma^n(\theta))^2), 
    \end{split}
\end{equation}
%$c^{ac}(\theta) | c^{ac}(\theta_{1:L^0}) \sim \mathcal{N}(\mu^n(\theta), (\Sigma^n)^2(\theta))$, 
where 
$\mu^n(\theta)=\Sigma^0 (\theta,\theta_{1:l})\cdot \Sigma^0 (\theta_{1:l},\theta_{1:l})^{-1} \cdot (c^{ac}(\theta_{1:l})-\mu^0(\theta_{1:l}))+\mu^0(\theta)$ 
and
$(\Sigma^n(\theta))^2= \Sigma^0 (\theta,\theta)-\Sigma^0 (\theta,\theta_{1:l})\cdot \Sigma^0 (\theta,\theta_{1:l})^{-1} \cdot \Sigma^0 (\theta_{1:l},\theta) $. 

We adopt \textit{expected improvement} as the acquisition function. 
Define $c^*_l:=\max_{l'\in \{1,\cdots,l\} } c^{ac}(\theta_{l'})$ as the optimal evaluation among the first $l$ evaluations and a shorthand notation $(c^{ac}(\theta)-c^*_l)^+ := \max\{c^{ac}(\theta)-c^*_l, 0\}$. 
For any $l\in \mathcal{L}$, we define $\mathbb{E}_l[\cdot]:=\mathbb{E}[\cdot|c^{ac}(\theta_{1:l})] $ as the expectation taken under the posterior distribution of $c^{ac}(\theta)$ conditioned on the values of $l$ evaluations $c^{ac}(\theta_{1:l})$. 
Then, the expected improvement is $\text{EI}_l(\theta):=\mathbb{E}_l [ (c^{ac}(\theta)-c^*_l)^+ ]$. 
The hyperparameter at the next tuning stage is chosen to maximize the expected improvement at the current stage, i.e, 
\begin{equation}
\label{eq:EI}
    \theta_{l+1}\in \textrm{arg}\max_{\theta\in \Theta^d} \text{EI}_l(\theta). 
\end{equation}
The expected improvement can be evaluated in a closed form, and \eqref{eq:EI} can be computed inexpensively by gradient methods  \cite{frazier2018bayesian}. 

At the first $L^0\in \{1,2,\cdots,L\}$ tuning stages, we choose the hyperparameter $\theta_l, l\in \{1,2,\cdots,L^0\}$, uniformly from $\Theta^d$. 
We can use the evaluation results $c^{ac}(\theta_l), l\in \{1,2,\cdots,L^0\}$,  to determine the parameters $\bar{\mu}^0, \lambda^0$, and $\lambda^i, i\in \{1,2,\cdots,d\}$, by Maximum Likelihood Estimation (MLE); i.e., we determine the values of these parameters so that they maximize the likelihood of observing the vector $[c^{ac}(\theta_{1:L^0})]$.  
For the remaining $L-L^0$ tuning stages, we choose $\theta_l, l\in \{L^0,L^0+1,\cdots,L\}$, in sequence as summarized in Algorithm \ref{algorithm:BO}. 

\begin{algorithm}[h]
\SetAlgoLined
%\small 
%\footnotesize
%\textbf{Input:} Number of trials $L$, Number of initial sample $L^0$\;
  \textbf{Implement} the initial $L^0$ evaluations $c^{ac}(\theta_l), l\in \{1,2,\cdots,L^0\}$\;
  %Gaussian process prior over function $c$\; 
  \textbf{Place} a Gaussian process prior on $c^{ac}\in \mathcal{C}$, i.e.,  $c^{ac}(\theta_{1:L^0})\sim \mathcal{N}(\mu^0(\theta_{1:L^0}),\Sigma^0(\theta_{1:L^0},\theta_{1:L^0}))$\;
\For{$l \leftarrow L^0$ \KwTo $L$}{
\textbf{Obtain} the posterior distribution of $c^{ac}(\theta)$ in \eqref{eq:conditionDis} based on the existing $l$ evaluations\;
\textbf{Compute} $\text{EI}_l(\theta), \forall \theta\in \Theta^d$, based on the posterior distribution\;
\textbf{Determine} $\theta_{l+1}$ via \eqref{eq:EI}\; 
\textbf{Implement} $\theta_{l+1}$ at the next tuning stage $l+1$ to evaluate $c^{ac}(\theta_{l+1})$\;
  }
  \textbf{Return} the maximized value of all observed samples, i.e., $\theta^*\in \textrm{arg}\max_{\theta_l\in \{\theta_1,\cdots,\theta_L\}} c^{ac}(\theta_l)$\; %at the final tuning stage as the optimal hyperparameter\; 
 \caption{Hyperparameter tuning via BO.\label{algorithm:BO}}
\end{algorithm}

\section{Case Study}
\label{sec:case study}

In this case study, we verify the effectiveness of ADVERT via a data set collected from human subject experiments conducted at New York University \cite{cox2020stuck}. 
We elaborate on the experiment setup and the data processing procedure in Section \ref{sec:experimentSetting}. 
Based on the features obtained from the data set, we generate synthetic data under adaptive visual aids to demonstrate the proposed attention enhancement mechanism and the phishing prevention mechanism in Section \ref{sec:validation of attention Enhance} and \ref{sec:validation of phishing prevention}, respectively. 
%The synthetic data replace an agent-based simulation which would be our future work. 

\subsection{Experiment Setting and Data Processing}
\label{sec:experimentSetting}
%The experiment involves $M=200$ undergraduate students\footnote{Due to low sampling rates and failures in eye-tracking calibration, $40$ of them were removed from analyses.} 
The data set involves $M=160$ undergraduate students ($n_{\text{White}}=27$, $n_{\text{Black}}=19$, $n_{\text{Asian}}=64$, $n_{\text{Hispanic/Latinx}}=17$, $n_{\text{other}}=33$) who are asked to vet $N=12$ different emails (e.g., the email of NYU friends network in Fig. \ref{fig:sampleEmail}) separately and then give a rating of how likely they would take actions solicited in the emails (e.g., maintain membership in Fig. \ref{fig:sampleEmail}). 
When presented to different participants, each email is described as either posing a cyber threat or risk-free legitimate opportunities to investigate how the above description affects the participants' phishing recognition. 

\begin{figure}[h]
\centering
\includegraphics[width=0.48 \textwidth]{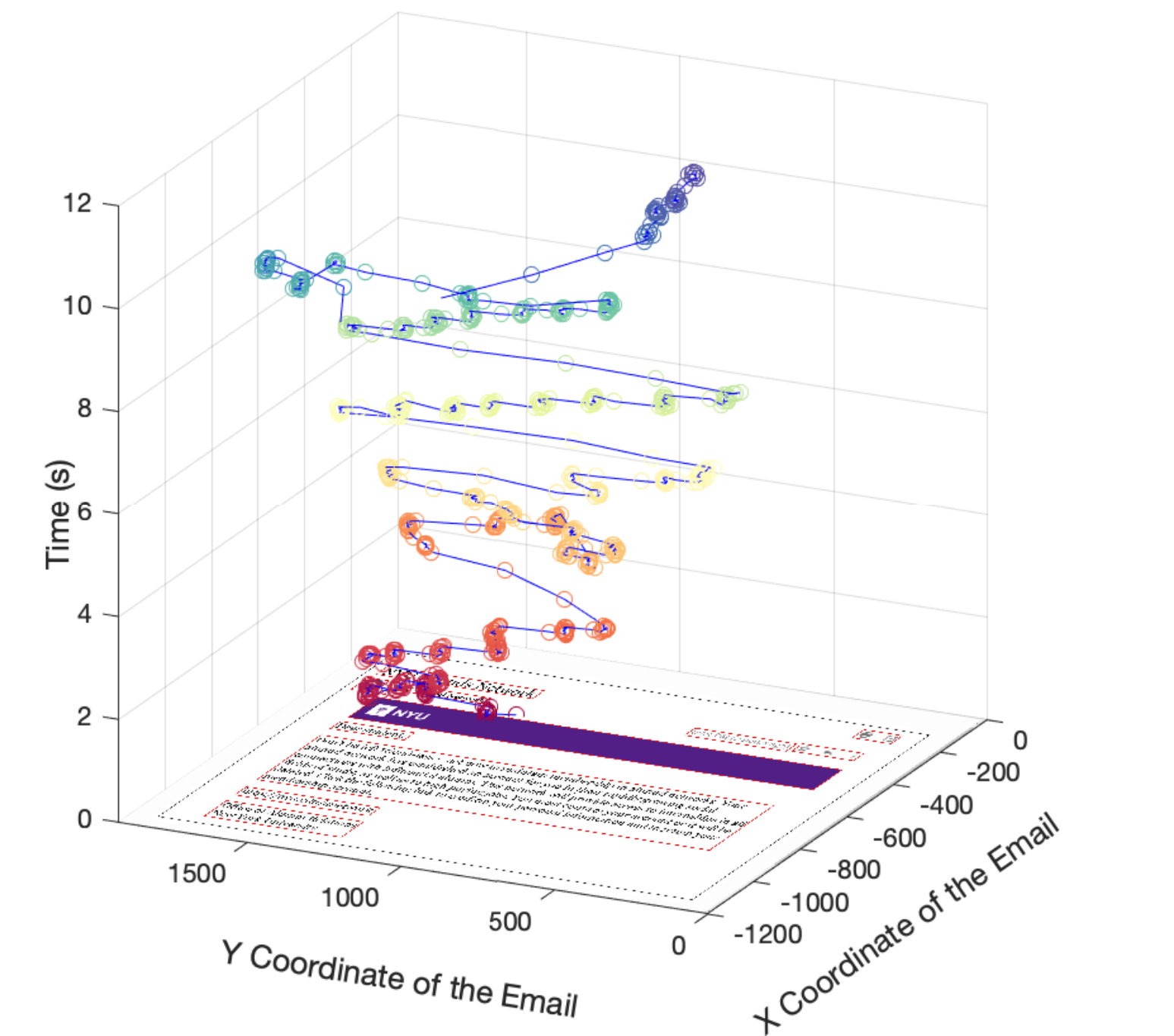}
\caption{ 
\textcolor{black}{A time-expanded plot of a typical eye-gaze trajectory with a sampling rate of $60$ Hz. 
The $x$-$y$ plane (in the unit of pixels) represents the email area. 
The $z$-axis represents the time (in the unit of seconds) of the participant's eye-gaze trajectory. 
The warmer color indicates a smaller value on the $z$-axis (i.e., an earlier gaze of the point).} 
}
\label{fig:3Deye_gaze}
\end{figure}

While the participants vet the emails, the Tobii Pro T60XL eye-tracking monitor records their eye locations on a $1920\times 1200$ resolution screen and the current pupil diameters of both eyes with a sampling rate of $60 \text{Hz}$. 
\textcolor{black}{Fig. \ref{fig:3Deye_gaze} illustrates the time-expanded eye-gaze trajectory of a participant vetting the sample email in Fig. \ref{fig:sampleEmail}. The $z$-coordinate of a 3D point $(x,y,z)$ represents the time when the participant gazes at the pixel $(x,y)$ in the email area. 
The participant's eye gaze locations move progressively from the points in warmer color to the ones in cooler color. 
Fig. \ref{fig:3Deye_gaze} illustrates the zigzag pattern of the participant's eye-gaze trajectory; i.e., the participant reads emails from left to right and top to bottom. 
The participant starts with the title, spends the majority of time on the main content, and glances at other AoIs (e.g., the links and the signatures). 
There is also a small chance of revisiting the email content and looking outside the email area.}

\begin{figure*}[h]
\centering
\includegraphics[width=1\textwidth]{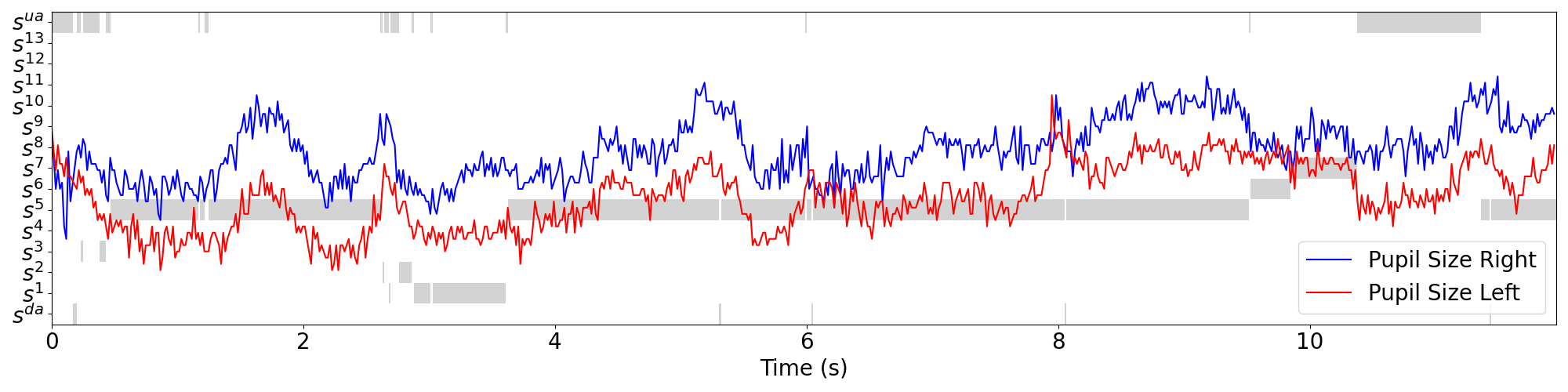}
\caption{
Gaze locations and pupil sizes collected in the trial of the data set \textcolor{black}{illustrated in Fig. \ref{fig:3Deye_gaze}}. 
The grey squares illustrate the transition of $15$ \textcolor{black}{VSs}. The red and blue lines represent the variations of the participant's left and right pupil sizes, respectively, as he reads the email. The $x$-axis represents the time \textcolor{black}{(in the unit of seconds) during} the email inspection.  
}
\label{fig:pupilsize}
\end{figure*}

Fig. \ref{fig:pupilsize} illustrates the participant's pupil sizes of left and right eyes in red and blue, respectively, \textcolor{black}{concerning the same trial of the data set to generate Fig. \ref{fig:3Deye_gaze}.} 
At different times, the average of the pupil diameters (resp. gaze locations) of the right and left eyes represent the pupil size (resp. gaze location). 
\textcolor{black}{Following Section \ref{sec:VS transition model}, we obtain the $15$ \textcolor{black}{VSs} illustrated by the grey squares in Fig. \ref{fig:pupilsize} based on the gaze locations of the email pixels in Fig. \ref{fig:3Deye_gaze}.} 
Since the covert eye-tracking system does not require head-mounted equipment or chinrests, the tracking can occur without the participants’ awareness. 
We refer the reader to the supplement materials of \cite{cox2020stuck} for the survey data and the details of the experimental procedure\footnote{\textcolor{black}{The processed data used in this manuscript, including the temporal transitions of AoIs and the pupil sizes, is available at 
\url{https://osf.io/4y32d/}. The raw eye-tracking data in the format of videos are available upon request.}}.

\subsubsection{Estimate Concentration Scores and Decay Rates based on Pupil Sizes}
\label{sec:attentionScorePupilSize}
Empirical works in \cite{janisse1973pupil,kang2014pupil} have demonstrated that pupils dilate as a consequence of attentional efforts. 
Building on the findings, we assume that the average pupil diameters of both eyes at time $t$ of the generation stage $k\in \mathcal{K}_m^n$ is approximately proportional to the participant's attention level $\frac{dv_k}{dt}(t)$ at time $t$.  % denoted by $\chi_k(t)$
We obtain the benchmark values of $r^{co}(s), \alpha(s), \forall s\in \mathcal{S}$, in Table \ref{table:AoIscore} by minimizing the \textcolor{black}{Mean Square Error (MSE)} between the CAL in Section \ref{sec:attentionEva} and the cumulative pupil size through global optimization methods such as Simulated Annealing (SA) \cite{van1987simulated}. %, i.e.,
% \begin{equation}
%     \min_{d} v_k(t)-\int_0^{}
% \end{equation}
%sample frequency is 60Hz
The results in Table \ref{table:AoIscore} corroborate that the main content AoI $s^5\in \mathcal{S}$ has the highest concentration score and the lowest decay rate. 
%The salutation, title is of second most important
\begin{table}[]
\centering
\begin{tabular}{|c|c|c|c|}
\hline
AoIs & Meaning        & $r^{co}(s^i)$ & $\alpha(s^i)$ \\ \hline
%0     & no data     & 18.17               & 10.81                       \\ \hline
$s^1$    & Title       & 9.48                & 2.17                        \\ \hline
$s^2$     & Sender      & 3.55                & 4.04                        \\ \hline
$s^3$     & Receiver    & 7.62                & 0.22                        \\ \hline
$s^4$     & Salutation        & 13.76               & 0.57                        \\ \hline
$s^5$      & Main Content     & 21.05               & 0.16                        \\ \hline
$s^6$      & URL         & 7.84                & 10.90                       \\ \hline
$s^7$      & Signature        & 6.47                & 5.46                        \\ \hline
$s^8$      & Logo        & 6.44                & 5.16                        \\ \hline
$s^9$      & Print\& Share    & 4.86                & 13.91                       \\ \hline
$s^{10}$     & Time        & 3.81                & 6.68                        \\ \hline
$s^{11}$    & Bookmark\& Forward      & 7.34                & 2.19                        \\ \hline
$s^{12}$    & Profile     & 7.26                & 2.02                        \\ \hline
$s^{13}$   & Attachment  & 4.74                & 3.46                        \\ \hline
%14    & other areas & 12.39               & 6.25                        \\ \hline
\end{tabular}
\caption{
The concentration score $r^{co}(s^i)$ and decay rate $\alpha(s^i)$ for $I=13$ AoIs. 
%Learning DAoI Score by Simulated Annealing. 
\label{table:AoIscore}}
\end{table}

\subsubsection{Synthetic VS Trajectory Generation under Visual Aids}
\label{sec:semi-markov}
In the case study, we consider $I=13$ AoIs. The sample email in Fig. \ref{fig:sampleEmail} illustrates the first $12$ AoIs. The $13$-th AoI is on the email attachment. 
%each participant's raw eye-tracking data to 12 segments, each of them corresponding to one trial, which starts from the moment a new email appears on the screen and ends when that email fades away and make decision. We extract below features for each trial:
\textcolor{black}{Under} visual aid $a\in\mathcal{A}$, we denote $P^{i,j}(a)$ as the probability of attention arriving at \textcolor{black}{VS $s^j\in \mathcal{S}$} from \textcolor{black}{VS $s^i\in \mathcal{S}$} and $\phi^i(a)$ as the average sojourn time at \textcolor{black}{VS $s^i\in \mathcal{S}$}. 
We specify the participants' VS transition trajectory $[s_t]_{t\in [0,T_m^n]}, \forall m\in \mathcal{M},n\in \mathcal{N}$, %under visual-aid generation policy $\sigma\in \Sigma$ 
as a semi-Markov transition process with probability transition matrix \textcolor{black}{ $P(a):=[P^{i,j}(a)]_{s^i,s^j\in \mathcal{S}}$ and exponential sojourn distribution of the scale parameter $\phi(a):=[\phi^i(a)]_{s^i\in \mathcal{S}}, \forall a\in \mathcal{A}$}. 

In particular, we consider a binary set of visual aid $\mathcal{A}=\{a^N, a^Y\}$, where $a^N$ represents the benchmark case without visual aids and $a^Y$ represents the visual aid of highlighting the entire email contents. 
%the entire email area. 
Based on the VS transition trajectory from the data set, we obtain the probability transition matrix $P(a^N)$ and the sojourn distribution parameter $\phi(a^N)$ under the benchmark case $a^N$. 
The transition matrix $P(a^Y)$ and sojourn distribution $\phi(a^Y)$ under visual aid $a^Y$ modify $P(a^N)$ and $\phi(a^N)$ based on the following observations. 
%Then, we revise the benchmark $P(a^N)$ and $\phi(a^N)$ to obtain $P(a^Y)$ and $\phi(a^Y)$ under visual aid $a^Y$. 
%some psych reference to support our assumption? 
On the one hand, the visual aid $a^Y$ decreases \textcolor{black}{$P^{i,{ua}}(a^Y), P^{i,{da}}(a^Y), \forall s^i\in \mathcal{S}$}; i.e., the participants will be guided by the visual aid to pay more frequent attention to the AoIs than the uninformative and distraction areas. 
On the other hand, the visual aid $a^Y$ decreases \textcolor{black}{$\phi^5(a^Y)$}; i.e., the persistent highlighting makes participants weary and reduces their attention spans on the email's main content. 

\begin{figure}[ht]
     \centering
     \begin{subfigure}[b]{0.24\textwidth}
         \centering
         \includegraphics[width=\textwidth]{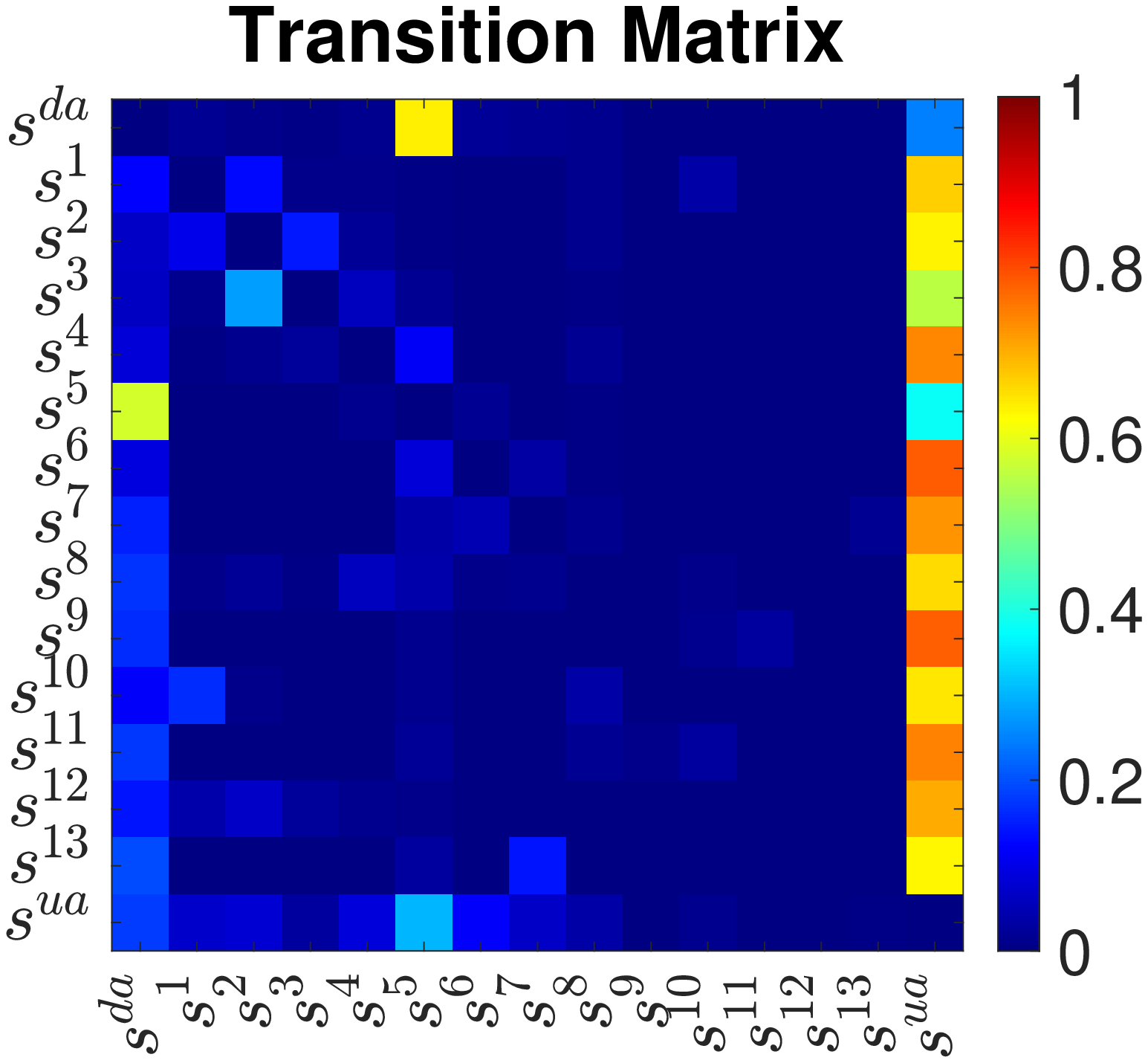}
         %\caption{Legit 0 (scam)}
         \caption{Under visual aid $a^N$.}
         \label{fig:heatmap_leg0}
     \end{subfigure}
     \hfill
     \begin{subfigure}[b]{0.24\textwidth}
         \centering
         \includegraphics[width=\textwidth]{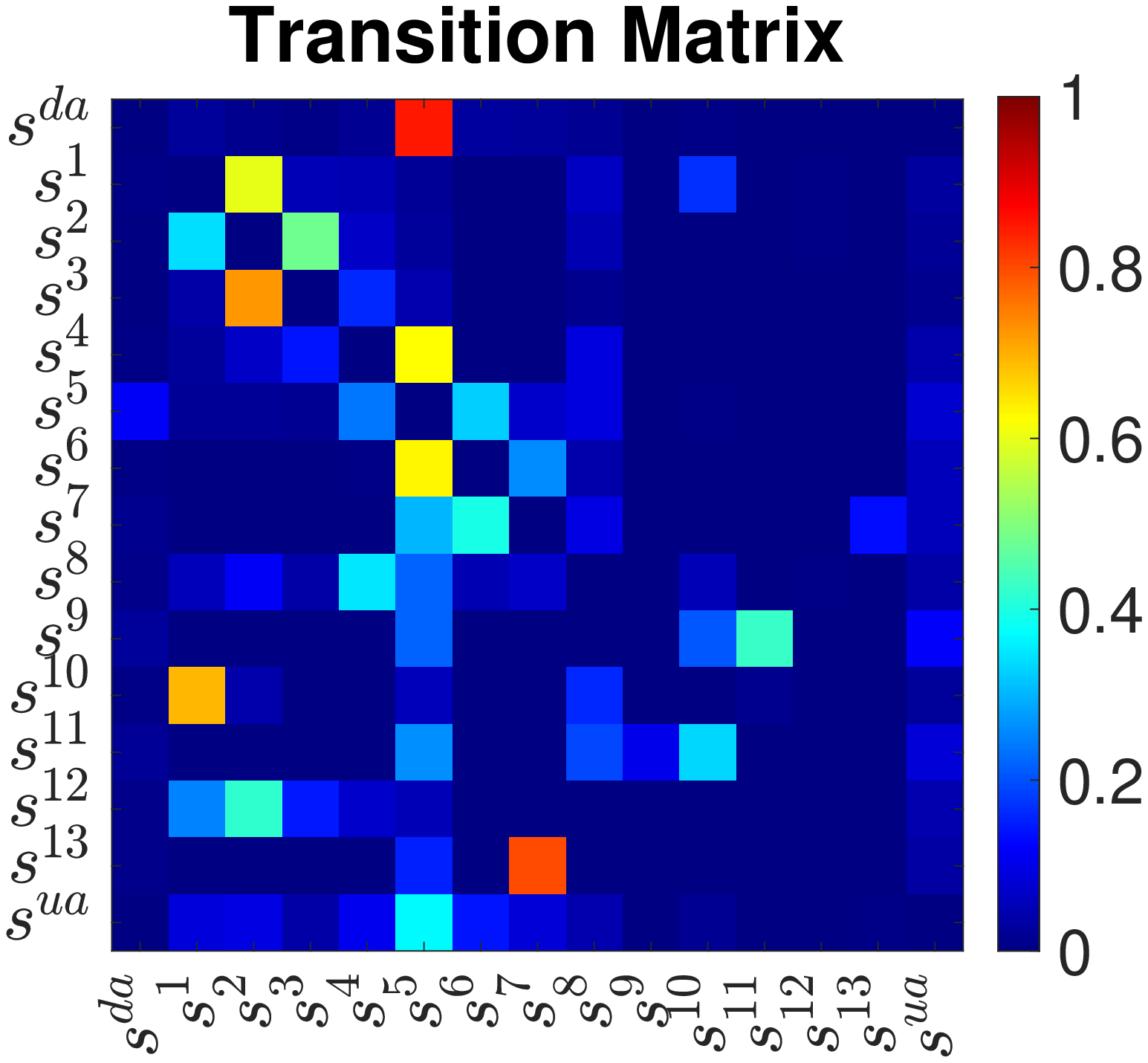}
         %\caption{Legit 1 (legitimate)}
          \caption{Under visual aid $a^Y$.}
         \label{fig:heatmap_leg1}
     \end{subfigure}
        \caption{
        Heat maps of the transition matrices $P(a),a\in \mathcal{A}$.  
        The row and the column represent the source and the destination of the $I+2$ \textcolor{black}{VSs}, respectively.  
        Under $a^Y$, the participants tend to pay attention to AoIs rather than the uninformative and distraction areas. 
        }
        \label{fig:heatmap}
\end{figure}

\begin{figure}[h]
\centering
\includegraphics[width=1\columnwidth]{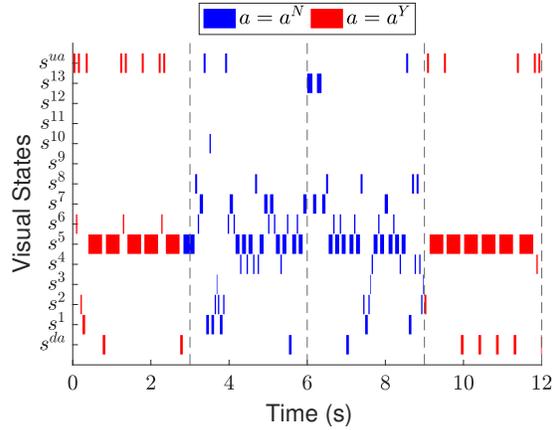}
\caption{
The VS transition trajectory when the visual aids in four generation stages are $a^Y,a^N,a^N$, and $a^Y$, respectively. 
The inspection lasts for $12$ seconds and the period length $T^{pl}$ is $3$ seconds. 
}
\label{fig:AOIpath}
\end{figure}

We illustrate $P(a^N)$ and $P(a^Y)$ using heat maps in Fig. \ref{fig:heatmap_leg0} and Fig. \ref{fig:heatmap_leg1}, respectively. 
In Fig. \ref{fig:AOIpath}, we illustrate an exemplary transition trajectory of $I+2$ \textcolor{black}{VSs} under $a^N$ and $a^Y$ in blue and red, respectively. 
The trajectory corroborates that participants under visual aid $a^Y$ incline to pay attention to AoIs yet have less sustained attention. 
\textcolor{black}{Accurately quantifying} the impact of the visual aid on the VS transition depends on many factors \cite{holland2013complex}, including the graphic design, the human subject, and the cognitive task. 
\textcolor{black}{In Section \ref{sec:semi-markov}}, we provide one potential estimation of the impact based on the human experiments to illustrate the implementation procedure and the effectiveness of the ADVERT framework. 

%%NOTE that in our case, we cannot reach t=infinity%%
% Based on $P(a),\phi(a), a\in \{a^N,a^Y\}$, we can compute the limiting occupancy distributions of the semi-Markov process \cite{nakagawa2011stochastic,huang2019adaptive} under $a^N$ and $a^Y$ in red and blue, respectively, as shown in Fig. \ref{fig:stationaryDist}. 
% This plot illustrates the comprehensive impact of visual aid on the visual state generation. %; i.e., when visual aid $a^Y$ is applied, at any time 
% \begin{figure}[h]
% \centering
% \includegraphics[width=1\columnwidth]{newfig/stationaryDist.eps}
% \caption{
% Limiting occupancy distribution of $I+2$ visual states under $a^N$ and $a^Y$ in red and blue, respectively. For each visual state $s\in \mathcal{S}$ in the x-axis, its y-value represents the probability of $s_t=s$ as $t\rightarrow \infty$. 
% }
% \label{fig:stationaryDist}
% \end{figure}

\subsection{Validation of Attention Enhancement Mechanism}
\label{sec:validation of attention Enhance}
\begin{figure}[h]
\centering
\includegraphics[width=1\columnwidth]{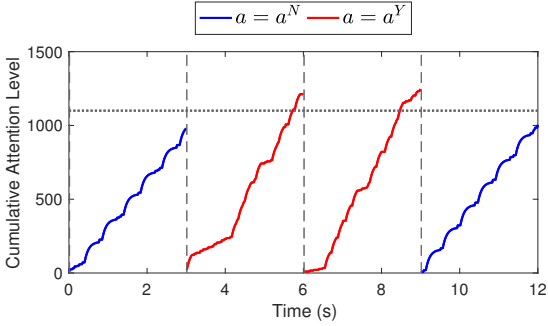} 
\caption{
The CAL of the  VS transition trajectory shown in Fig. \ref{fig:AOIpath}. 
The horizontal dotted line represents the attention threshold $X^{at}$. 
The visual aids in four generation stages are  $a^Y,a^N,a^N$, and $a^Y$, respectively, and
the resulting attention states are $x^L,x^H,x^H$, and $x^L$, respectively. 
}
\label{fig:AttentionScorePath}
\end{figure}

\begin{figure}[h]
\centering
\includegraphics[width=1\columnwidth]{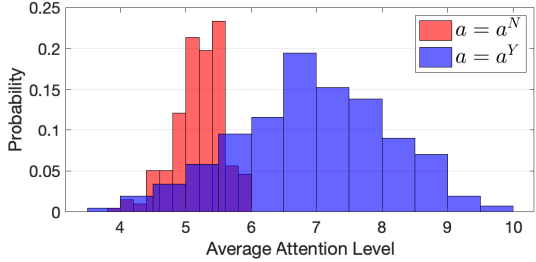}
\caption{
The normalized histogram of average attention level under visual aids $a^N$ and $a^Y$ in red and blue, respectively. 
}
\label{fig:hisAAL}
\end{figure}

Based on the benchmark attention score in Section \ref{sec:attentionScorePupilSize}, Fig. \ref{fig:AttentionScorePath} illustrates the CAL of the  VS transition trajectory shown in Fig. \ref{fig:AOIpath}. 
We consider $X=2$ attention states $\mathcal{X}=\{x^H,x^L\}$ with the \textit{attentive state} $x^H$ and the \textit{inattentive state} $x^L$. 
Define $X^{at}\in \mathbb{R}$ as the \textit{attention threshold}. 
If the AAL at generation stage $k\in \mathcal{K}_m^n$ is higher (resp. lower) than the attention threshold, i.e., $\bar{v}_k\geq X^{at}$ (resp. $\bar{v}_k\leq X^{at}$), then the attention state $x_k\in \mathcal{X}$ at generation $k$ is the attentive state $x^H$ (resp. inattentive state $x^L$). 
Fig. \ref{fig:hisAAL} further illustrates the impact of visual aids $a^N$ and $a^Y$ on the AAL in red and blue, respectively. 
The figure demonstrates that $a^Y$ can increase the mean of AAL yet increase its variance.   

In Algorithm \ref{algorithm:perEmail}, we present the Q-learning process for participant $m\in \mathcal{M}$ who \textcolor{black}{reads} email $n\in \mathcal{N}$ for $T_m^n$ seconds. 
Define $\eta_k(x,a)$ as the total number of visits to attention state $x\in \mathcal{X}$ and visual aid $a\in \mathcal{A}$ up to generation stage $k$.  
Then, we choose the learning rate $\gamma_k(x_k,a_k)=\frac{\eta^0}{\eta_k(x,a)-1+\eta^0}$ for all $x_k\in\mathcal{X},a_k\in \mathcal{A}$ to guarantee the asymptotic convergence, where $\eta^0\in(0,\infty)$ is a constant parameter. 

% a total duration of $\sum_{m\in \mathcal{M}}\sum_{n\in \mathcal{N}} T_m^n$ 

Based on the benchmark data set of $M=160$ participants who inspect $N=12$ emails in Section \ref{sec:experimentSetting}, the inspection time $T_m^n, \forall m\in \mathbf{M},n\in \mathcal{N}$, follows a \textit{Burr distribution}; i.e., its cumulative distribution function is described by $F^{Burr}(t \mid \rho_1, \rho_2, \rho_3)=1-\frac{1}{\left(1+\left({t}/{\rho_1}\right)^{\rho_2}\right)^{\rho_3}}$ with the scale parameter $\rho_1 = 11.7$, and the shape parameters $\rho_2=62.5,\rho_3=0.04$. The average inspection time of $M\times N$ samples is $18.7$ seconds. 
% We fit the time of diff people read diff emails by burr distribution with pdf
% \begin{align}
%     f(x \mid \alpha, c, k)=\frac{\frac{k c}{\alpha}\left(\frac{x}{\alpha}\right)^{c-1}}{\left(1+\left(\frac{x}{\alpha}\right)^{c}\right)^{k+1}}, \quad x>0, \alpha>0, c>0, k>0, 
% \end{align}
% and cdf
% \begin{equation}
%     F(x \mid \alpha, c, k)=1-\frac{1}{\left(1+\left(\frac{x}{\alpha}\right)^{c}\right)^{k}}, \quad x>0, \alpha>0, c>0, k>0,  
% \end{equation}
% where $c$ and $k$ are the shape parameters and $\alpha$ is the scale parameter. 
% \begin{figure}[h]
% \centering
% \includegraphics[width=1\columnwidth]{fig/InspTimeFit.pdf}
% \caption{
% Fit it with burr distribution with parameters $\alpha = 11700.2$, $c=62.5387$, and $k=0.0388$. The unit of data is ms. %with sample frequency 60HZ. 
% The inspection time is  $18.7109s$ on average. 
% }
% \label{fig:InspTimeFit}
% \end{figure}
During $T_m^n$ seconds of the email vetting process, the eye-tracking device records the participant's gaze locations, which leads to the VS transition trajectory. 
In Algorithm \ref{algorithm:perEmail}, we simulate the human email-reading process through the synthetic VS transition trajectory generated by the sufficient statistics $P(a_t)$ and $\phi(a_t)$. 
Every $T^{pl}$ seconds, ADVERT updates the Q-matrix and the visual aid based on \eqref{eq:Qlearning}. 

\begin{algorithm}[h]
\SetAlgoLined
%\small 
%\footnotesize
\textbf{Input:} Initial Q-matrix $[Q_0(x,a)]_{x\in \mathcal{X},a\in \mathcal{A}}$, initial attention state $x_0\in \mathcal{X}$, %the attention scores $r^{co}(s),\alpha(s), \forall s\in \mathcal{S}$, 
the number of visits $\eta_k(x,a)$, and the hyperparameter $\theta=[X^{at}, T^{pl}]$\; %the initial $\epsilon_0\in [0,1]$\;
  \textbf{Initialize} time $t=0$ and the inspection length $T_m^n$ based on the Burr distribution $F^{Burr}$\; 
  \textbf{Set} the initial visual aid $a_0\in \mathcal{A}$ based on the initial Q-matrix $Q_0$, the initial attention state $x_0$ and the $\epsilon_k$-greedy policy in Section \ref{sec:Q-learning}\;
\While{$t<T_m^n$}{
\textbf{Obtain} VS transition $s_t\in \mathcal{S}$ based on $P(a_t)$ and $\phi(a_t)$ (i.e., use synthetic visual data to achieve Step $2$ \textcolor{black}{in} Fig. \ref{fig:assistiveSecurityDiag})\;
\textbf{Evaluate} the CAL $v_k(t)$ based on $r^{co},\alpha$  as shown \textcolor{black}{by} Step $3$ \textcolor{black}{in} Fig. \ref{fig:assistiveSecurityDiag}\;
\If{$t=kT^{pl}, k\in \mathbb{Z}^{+}$}
{
\leIf{$\bar{v}_k\geq X^{at}$ (shown \textcolor{black}{by} Step $4$ \textcolor{black}{in} Fig. \ref{fig:assistiveSecurityDiag})}{attentive attention state $x_k=x^H$}{inattentive attention state $x_k=x^L$} 
\textbf{Update} Q-matrix $Q_k$ based on \eqref{eq:Qlearning} as shown \textcolor{black}{by} Step $5$ \textcolor{black}{in} Fig. \ref{fig:assistiveSecurityDiag}\;
\textbf{Implement} the visual aid $a_k\in \mathcal{A}$ based on the current Q-matrix $Q_k$ and the $\epsilon_k$-greedy policy (i.e., Step $6$ \textcolor{black}{in} Fig. \ref{fig:assistiveSecurityDiag})\; 
\lIf{$x_k=x,a_k=a$}{update the number of visits $\eta_{k+1}(x,a)\leftarrow \eta_k(x,a)+1$}
\textbf{Output} the number of updates $K_m^n\leftarrow k$\;
  }
}
\textbf{Implement} the pre-trained neural network in Section \ref{sec:NN} to estimate whether participant $m$ has made the correct judgment concerning email $n$, i.e., $z_m^n(\theta)\in \{z^{co},z^{wr}\}$ (i.e., use synthetic decision data to achieve Step $7$ \textcolor{black}{in} Fig. \ref{fig:assistiveSecurityDiag})\;
\textbf{Return:} Q-matrix $[Q_{K_m^n}(x,a)]_{x\in \mathcal{X},a\in \mathcal{A}}$, final attention state $x_{K_m^n}\in \mathcal{X}$, number of visits $\eta_{K_m^n}(x,a)$, and $z_m^n(\theta)$\;

 \caption{[\textbf{Individual Adaptation}] Optimal visual-aid learning and attention enhancement for participant $m\in \mathcal{M}$ vetting email $n\in \mathcal{N}$. 
 \label{algorithm:perEmail}}
\end{algorithm}

Following Section \ref{sec:Q-learning}, we develop Algorithm \ref{algorithm:attentionEnhance} to illustrate the entire attention enhancement loop that involves the consolidation of the data set from $\bar{M}\in \{1,\cdots,M\}$ participants and $\bar{N}\in \{1,\cdots,N\}$ emails. 
%In the loop (lines $28$ to $32$)
After the participant $m\in  \{1,\cdots,\bar{M}\}$ finishes reading the email $n\in  \{1,\cdots,\bar{N}\}$, Algorithm \ref{algorithm:perEmail} returns the Q-matrix and the attention state at the final generation stage $K_m^n$. 
These results then serve as the inputs for the next email inspection until $N^{bo}$ emails have been inspected. 
\begin{algorithm}[h]
\SetAlgoLined
%\small 
%\footnotesize
 \caption{[\textbf{Population Adaptation}] Optimal visual-aid learning through a consolidated data set of  $\bar{M}\in \{1,\cdots,M\}$ participants vetting $\bar{N}\in \{1,\cdots,N\}$ emails.  
 \label{algorithm:attentionEnhance}}
\textbf{Input:}  Hyperparameter $\theta=[X^{at}, T^{pl}]$\;
  \textbf{Initialize} Q-matrix $[Q_0(x,a)]_{x\in \mathcal{X},a\in \mathcal{A}}$ as a zero matrix, $\eta_0(x,a)=0, \forall x\in \mathcal{X},a\in \mathcal{A}$, and initial attention state $x_0\in \mathcal{X}$\;
\For{participant $m \in \{1,\cdots,\bar{M}\}$ vetting email $n\in \{1,\cdots,\bar{N}\}$}
{
\textbf{Implement} Algorithm \ref{algorithm:perEmail} with the inputs of $[Q_0(x,a)]_{x\in \mathcal{X},a\in \mathcal{A}}$, $x_0\in \mathcal{X}$, and $\eta_0(x,a)$\;
\textbf{Save} the outputs of $[Q_{K_m^n}(x,a)]_{x\in \mathcal{X},a\in \mathcal{A}}$, $x_{K_m^n}\in \mathcal{X}$, $\eta_{K_m^n}(x,a)$, and $z_m^n(\theta)$\;
\textbf{Cascade} the outputs to the inputs of the next loop: $Q_0 \leftarrow Q_{K_m^n}$, $x_0\leftarrow x_{K_m^n}$, and $\eta_{0}\leftarrow \eta_{K_m^n}$\; 
}
\textbf{Return}: the accuracy metric  $c^{ac}(\theta)$ based on \eqref{eq:accuracy metric}\;
\end{algorithm}

\begin{figure}[h]
\centering
\includegraphics[width=1\columnwidth]{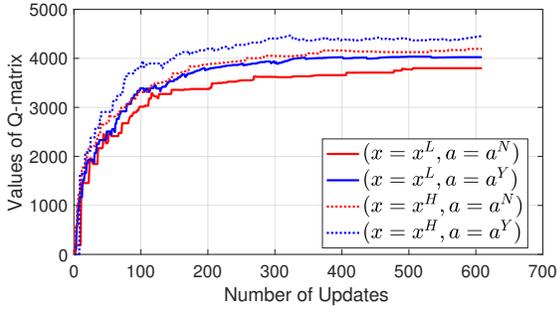}
\caption{
The Q-learning updates under hyperparameters $X^{at}=5.56$ and $T^{pl}=3$ seconds. 
The red and blue lines represent the  Q-matrix values under visual aids $a^N$ and $a^Y$, respectively. 
The solid and dashed lines represent the  Q-matrix values under attention states $x^L$ and $x^H$, respectively. 
}
\label{fig:Qlearning}
\end{figure}

Based on Algorithm \ref{algorithm:attentionEnhance}, we plot the entire Q-learning updates with $N^{bo}=100$ emails in Fig. \ref{fig:Qlearning} that contains a total of $609$ generations stages. 
The learning results show that the visual aid $a^Y$ outweighs $a^N$ for both attention states and should be persistently applied under the current setting. 
%Thus, the visual aid $a^Y$ should be applied 

\subsection{Validation of Phishing Prevention Mechanism}
\label{sec:validation of phishing prevention}
After we obtain a participant's synthetic response (characterized by his VS transition trajectory) under the adaptive visual aids, we apply a pre-trained neural network to estimate whether the participant has made a correct judgment as shown in line $24$ of Algorithm \ref{algorithm:perEmail}. 
In Section \ref{sec:NN}, we elaborate on the training process of the neural network based on the data set used in Section \ref{sec:experimentSetting}. 
We apply \textcolor{black}{BO} in Algorithm \ref{algorithm:BO} to evaluate the accuracy metric  $c^{ac}\in \mathcal{C}$, as illustrated \textcolor{black}{by} Step $8$ \textcolor{black}{in} Fig. \ref{fig:assistiveSecurityDiag}. In Section \ref{sec:BOresults}, we show the results. 

\subsubsection{Neural Network}
\label{sec:NN}
%As stated in Section \ref{sec:experimentSetting}, our offline data set does not contain a ground-truth of whether an email is phishing or legitimate. 
In this case study, we regard the majority choice of the $M=160$ participants as the email's true label.  Without visual aids, these participants achieve an accuracy of $74.6\%$ on average. 
Under the assumption that the hyperparameters affect the participants' phishing recognition only through their VS transitions, we construct a neural network with an LSTM layer, a dropout layer, and a fully-connected layer to establish the relationship from the sequence of VS transition trajectory $[s_t]_{t\in T_m^n}$ to the label of judgment correctness $z_m^n\in \{z^{co},z^{wr}\}$. 
We split the entire trials of the \textcolor{black}{eye-tracking} data set into $1113$ training data and $128$ test data\footnote{\textcolor{black}{There are $1920$ trials in total, and we carefully exclude the remaining $679$ trials for two reasons. First, Tobii Pro T60XL records the participants' eye locations with a \textit{validity level} ranging from 0 (high confidence) to 4 (eye not found). We exclude a trial if more than $70\%$ of its vetting time has a validity value of $4$. 
Second, we exclude trials of irresponsible participants who spend the majority (i.e., over $70\%$) of time in uninformative areas. 
%\textcolor{black}{Third, some participants type on the keyboard more than necessary, which causes trouble to abstract the AoIs from the raw eye-tracking videos through an automated process. These trials removed for the third reason can be recovered by manual processing (or by recording the click of the button, we can do it, but we are just lazy. Also 155 is not too bad).}
}}. 
%Reason 3 correspond about 155 trials. Reason 1+2 correspond to about 524 trials.
The trained neural network achieves a sensitivity of $0.89$, a specificity of $0.21$, an f$1$-score of $0.73$, and an accuracy of $0.61$. 

\subsubsection{Bayesian Optimization Results}
\label{sec:BOresults}
\begin{figure}[h]
\centering
\includegraphics[width=1\columnwidth]{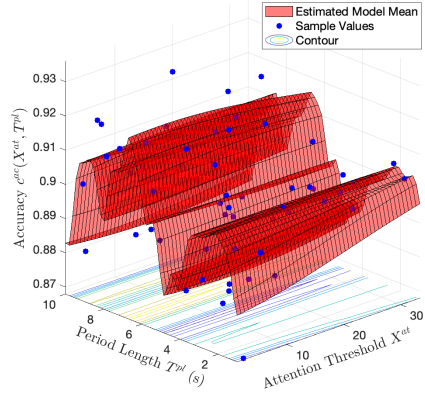} 
\caption{
The estimated Gaussian model of the objective function $c^{ac}(\theta)$ concerning the hyperparameter $\theta=[X^{at},T^{pl}]$ in red with its contour on the bottom. 
The blue points represent the sample values of $60$ \textcolor{black}{tuning stages}. 
}
\label{fig:BOmodel}
\end{figure}

As explained in Section \ref{sec:mata-learning}, for each different application scenario, a meta optimization of the accuracy metric $c^{ac}(X^{at},T^{pl})$ is required to find the optimal attention threshold $X^{at}$ and the period length $T^{pl}$ for visual-aid generation. 
To obtain the value of $c^{ac}(X^{at},T^{pl})$ under different values of the hyperparameter $\theta=[X^{at},T^{pl}]$, we need to implement the hyperparameter in Algorithm \ref{algorithm:attentionEnhance} and repeat for $n^{rp}$ times to reduce the noise. Thus, the evaluation is costly, and \textcolor{black}{BO} in Algorithm \ref{algorithm:BO} is a favorable method to achieve the meta optimization. 
We illustrate the \textcolor{black}{BO} for $L=60$ tuning stages in Fig. \ref{fig:BOmodel}. 
Each blue point represents the average value of $c^{ac}(X^{at},T^{pl})$ over $n^{rp}=20$ repeated samples under the hyperparameter $\theta=[X^{at},T^{pl}]$. 
Based on the estimated Gaussian model in red, we \textcolor{black}{observe} that the attention threshold $X^{at}\in [1,33]$ has a small impact on phishing recognition while the period length $T^{pl}\in [60,600]$ has a periodic impact on phishing recognition. 
The optimal hyperparameters for phishing prevention are $X^{at,*}=8.8347$ and $T^{pl,*}=6.63$ seconds. %398/60

\begin{figure}[h]
\centering
\includegraphics[width=1\columnwidth]{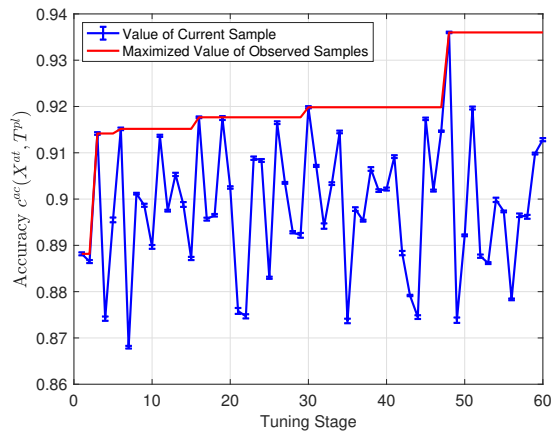} 
\caption{
Accuracy metric $c^{ac}(X^{at},T^{pl})$ at $L=60$ tuning stages. 
The blue line and its error bar represent the mean value of the samples and their variances, respectively. 
The red line represents the maximized value of the observed samples up to the current tuning stage. 
}
\label{fig:BOerror}
\end{figure}

We illustrate the temporal procedure of \textcolor{black}{BO for} $L=60$ tuning stages in Fig. \ref{fig:BOerror}. 
As we increase the number of tuning stages to obtain more samples, the maximized value of the accuracy metric $c^{ac}\in \mathcal{C}$ monotonously increases as shown in red. %with the number of tuning stages.  
The blue line and its error bar represent the mean and variances of the sample values at each tuning stage, respectively. 
Throughout the $L=60$ tuning stages, the variance remains small, which indicates that ADVERT is \textit{robust} to the noise of human attention and decision processes. 

Compared to the benchmark accuracy of $74.6\%$ without visual aids, participants with visual aid achieve the accuracy of a minimum of $86\%$ under all $60$ \textcolor{black}{tuning stages} of different hyperparameters. 
The above accuracy improvement corroborates that the ADVERT's attention enhancement mechanism \textcolor{black}{highlighted by the blue background in} Fig. \ref{fig:assistiveSecurityDiag} effectively serves as a stepping stone to facilitate phishing recognition. 
The results shown in the blue line further corroborate the efficiency of the ADVERT's phishing prevention mechanism \textcolor{black}{highlighted by the orange background in} Fig. \ref{fig:assistiveSecurityDiag}; i.e., in less than $50$ tuning stages, we manage to improve the accuracy of phishing recognition from $86.8\%$ to $93.7\%$. 
Besides, the largest accuracy improvement (from $88.7\%$ to $91.4\%$) happens within the first $3$ tuning stages. 
Thus, if we have to reduce the number of tuning stages due to budget limits, ADVERT can still achieve a sufficient improvement \textcolor{black}{in the accuracy of recognizing phishing}. 

\textcolor{black}{
\section{Limitations and Mitigation}
\label{sec:limitations}
The limitations of the data set and the data processing process are as follows. 
First, the demographic of the experimental subjects is limited to $160$ undergraduate students. 
%It can be expanded to provide sufficient data to enable a more comprehensive study of the human behaviors that cover different population groups. 
%, which may not be sufficiently representative to study human attention during the email vetting process. 
In the current work, we handle this issue by diversifying the participants (concerning their races, genders, and ages) and adopting the feedback loop of Bayesian optimization (that adapts to unconsidered user groups). 
%The feedback loop of Bayesian optimization also helps adapt the design to other unconsidered user groups. 
To enable a more comprehensive study of the human behaviors that cover different user groups, we can recruit more diversified participants through crowd-sourcing websites, including Amazon Mechanical Turk (MTurk).} 
\textcolor{black}{
Second, the dataset contains 12 unique emails. They are certainly not meant to be comprehensive to cover all phishing scenarios. %or transfer to other unseen data sets. 
However, they are sufficient for this work, which focuses on the system-level control of human attention processes to improve the accuracy of phishing recognition. For each email, we conduct the vetting processes of $M = 160$ humans, which result in the distinct $1241$ trials of eye-tracking trajectories. These eye-tracking trials are sufficient to reveal human attention patterns. Moreover, as a data-driven and system-level framework, ADVERT can adapt and generalize to unseen sets of emails.}
\textcolor{black}{
Third, we exclude approximately one-third of the eye-tracking data due to their low validity scores that arise from the limitation of the eye-tracking device and the imprudence of the participants, as stated in the footnote of Section \ref{sec:NN}.   
The reduced sample size may lead to overfitting issues. 
We can overcome it by improving the eye-tracking device, revising the experiment setting, and recruiting a sufficient number of participants.
%the manual processing of the raw eye-tracking video data. 
}

\section{Conclusions and Future Work}
\label{sec:conclusion}
As a prototypical \textit{innate human vulnerability}, lack of attention is one of the main challenges to protecting users from phishing attacks. 
To address the challenge, we have developed a \textit{human-technical solution} called ADVERT to guide the users' attention to the right contents of the email and consequently improve their accuracy \textcolor{black}{of} phishing recognition. 

To enable a real-time evaluation of the user's visual behaviors, we have built AoIs from the entire email area and a transition model to compress the eye-tracking data into a representative VS transition trajectory. 
After assigning the concentration rewards and decay parameters to evaluate the user's CAL, we have defined \textit{privacy-preserving} and \textit{light-weight} metrics, i.e., AAL and QAAL, to represent the user's attention state at each time of visual-aid generation. 
These metrics enable us to apply model-free \textcolor{black}{RL} methods and generate the optimal visual aid for real-time attention enhancement. 
Using the above attention enhancement mechanism as a stepping-stone, we have designed an efficient algorithm to tune the hyperparameters related to the visual aid generation pattern and the attention evaluation parameters. 
The update of these hyperparameters at each tuning stage revises the visual aids, affects the users' attention, and consequently improves the accuracy of phishing recognition. 

We have corroborated the effectiveness of ADVERT through a case study based on the data set collected from human subject experiments conducted at New York University. 
By abstracting the transition matrix and sojourn distribution from the data set as \textit{sufficient statistics} of the stochastic VS transition, we have generated synthetic VS transition to simulate the participant's visual behaviors under visual aids. 
Meanwhile, we have trained a neural network to estimate the correctness of the participant's phishing recognition based on the VS transition trajectory. 
%The concentration scores and decay rates estimated based on the participant's pupil size corroborate the significant roles of the main content AoI. 
Finally, we have developed two algorithms to design visual aids that adapt to each individual and the population, \textcolor{black}{respectively}. 
For the attention enhancement mechanism, the results have shown that the visual aids can statistically increase the AAL and improve the accuracy of phishing recognition from $74.6\%$ to a minimum of $86\%$. 
%The meta-adaptation loop can further increase the the accuracy of phishing recognition from  $86.8\%$ to  $93.7\%$ in less than $50$ tuning stages. 
The meta-adaptation has been shown to be \textit{effective} (e.g., improve the accuracy of phishing recognition from  $86.8\%$ to  $93.7\%$ in less than $50$ tuning stages), \textit{efficient} (e.g., the largest accuracy improvement happens within $3$ tuning stages), and \textit{robust} (e.g., the variances of $L=60$ samples remain small). 
The results have also provided insights and guidance for the ADVERT design; e.g., the attention threshold (resp. the period length) has a small (resp. periodic) impact on phishing recognition. 

The future work would focus on designing a more sophisticated visual support system that can determine when and how to generate visual aids in lieu of a periodic generation. 
We may also embed  ADVERT into VR/AR technologies to mitigate human vulnerabilities under simulated deception scenarios, where the simulated environment can be easily repeated or changed. 
%We would allow the participants to report their confidence levels while they make the security decisions. 
Finally, there would be an opportunity to incorporate factors such as pressure and incentives into the design by limiting the participant's vetting time and providing rewards for accurately identifying phishing, respectively. 

% % use section* for acknowledgment
\section*{Acknowledgment}
The authors would like to thank Jennie W. Qu-Lee and Blair Cox for their help to export and interpret the eye-tracking data set housed on the experiment platform of the NYU Social Perception Action \& Motivation (SPAM) laboratory. %led by Professor Emily Balcetis. 
\textcolor{black}{We thank Prof. Jonathan Bakdash and the other anonymous reviewer for the helpful comments on earlier drafts of the manuscript.}

%Bibliography 
\bibliographystyle{ieeetr}
\bibliography{sample}

\begin{thebibliography}{10}

\bibitem{aldawood2018educating}
H.~Aldawood and G.~Skinner, ``Educating and raising awareness on cyber security
  social engineering: A literature review,'' in {\em 2018 IEEE International
  Conference on Teaching, Assessment, and Learning for Engineering (TALE)},
  pp.~62--68, IEEE, 2018.

\bibitem{alotaibi2016information}
M.~Alotaibi, S.~Furnell, and N.~Clarke, ``Information security policies: A
  review of challenges and influencing factors,'' in {\em 2016 11th
  International Conference for Internet Technology and Secured Transactions
  (ICITST)}, pp.~352--358, IEEE, 2016.

\bibitem{huang2021duplicity}
L.~Huang and Q.~Zhu, ``Duplicity games for deception design with an application
  to insider threat mitigation,'' {\em IEEE Transactions on Information
  Forensics and Security}, vol.~16, pp.~4843--4856, 2021.

\bibitem{huang2022zetar}
L.~Huang and Q.~Zhu, ``Zetar: Modeling and computational design of strategic
  and adaptive compliance policies,'' {\em arXiv preprint arXiv:2204.02294},
  2022.

\bibitem{cox2020stuck}
E.~B. Cox, Q.~Zhu, and E.~Balcetis, ``Stuck on a phishing lure: differential
  use of base rates in self and social judgments of susceptibility to cyber
  risk,'' {\em Comprehensive Results in Social Psychology}, vol.~4, no.~1,
  pp.~25--52, 2020.

\bibitem{dhamija2006phishing}
R.~Dhamija, J.~D. Tygar, and M.~Hearst, ``Why phishing works,'' in {\em Proc.
  of the SIGCHI conference on Human Factors in computing systems},
  pp.~581--590, 2006.

\bibitem{WinNT1}
I.~Baxter, ``Fake login attack evades logo detection,'' 2020.
\newblock
  \url{https://ironscales.com/blog/fake-login-attack-evades-logo-detection}.

\bibitem{wen2019hack}
Z.~A. Wen, Z.~Lin, R.~Chen, and E.~Andersen, ``What. hack: engaging
  anti-phishing training through a role-playing phishing simulation game,'' in
  {\em Proceedings of the 2019 CHI Conference on Human Factors in Computing
  Systems}, pp.~1--12, 2019.

\bibitem{dodge2007phishing}
R.~C. Dodge~Jr, C.~Carver, and A.~J. Ferguson, ``Phishing for user security
  awareness,'' {\em computers \& security}, vol.~26, no.~1, pp.~73--80, 2007.

\bibitem{jain2017phishing}
A.~K. Jain and B.~B. Gupta, ``Phishing detection: analysis of visual similarity
  based approaches,'' {\em Security and Communication Networks}, vol.~2017,
  2017.

\bibitem{khonji2013phishing}
M.~Khonji, Y.~Iraqi, and A.~Jones, ``Phishing detection: a literature survey,''
  {\em IEEE Communications Surveys \& Tutorials}, vol.~15, no.~4,
  pp.~2091--2121, 2013.

\bibitem{kelley2016real}
T.~Kelley and B.~I. Bertenthal, ``Real-world decision making: Logging into
  secure vs. insecure websites,'' {\em Proceedings of the USEC}, vol.~16,
  no.~10.14722, 2016.

\bibitem{egelman2008you}
S.~Egelman, L.~F. Cranor, and J.~Hong, ``You've been warned: an empirical study
  of the effectiveness of web browser phishing warnings,'' in {\em Proceedings
  of the SIGCHI Conference on Human Factors in Computing Systems},
  pp.~1065--1074, 2008.

\bibitem{al2019autonomous}
E.~Al-Shaer, J.~Wei, W.~Kevin, and C.~Wang, ``Autonomous cyber deception,''
  {\em Springer}, 2019.

\bibitem{huang2020dynamic}
L.~Huang and Q.~Zhu, ``A dynamic games approach to proactive defense strategies
  against advanced persistent threats in cyber-physical systems,'' {\em
  Computers \& Security}, vol.~89, p.~101660, 2020.

\bibitem{pawlick2021game}
J.~Pawlick and Q.~Zhu, {\em Game Theory for Cyber Deception: From Theory to
  Applications}.
\newblock Springer Nature, 2021.

\bibitem{katsini2020role}
C.~Katsini, Y.~Abdrabou, G.~E. Raptis, M.~Khamis, and F.~Alt, ``The role of eye
  gaze in security and privacy applications: survey and future hci research
  directions,'' in {\em Proceedings of the 2020 CHI Conference on Human Factors
  in Computing Systems}, pp.~1--21, 2020.

\bibitem{ramkumar2020eyes}
N.~Ramkumar, V.~Kothari, C.~Mills, R.~Koppel, J.~Blythe, S.~Smith, and A.~L.
  Kun, ``Eyes on urls: Relating visual behavior to safety decisions,'' in {\em
  ACM Symposium on Eye Tracking Research and Applications}, pp.~1--10, 2020.

\bibitem{miyamoto2015eye}
D.~Miyamoto, G.~Blanc, and Y.~Kadobayashi, ``Eye can tell: On the correlation
  between eye movement and phishing identification,'' in {\em Int. Conf. on
  Neural Information Processing}, pp.~223--232, Springer, 2015.

\bibitem{mcalaney2020understanding}
J.~McAlaney and P.~J. Hills, ``Understanding phishing email processing and
  perceived trustworthiness through eye tracking,'' {\em Front. Psychol.},
  vol.~11, p.~1756, 2020.

\bibitem{xiong2017domain}
A.~Xiong, R.~W. Proctor, W.~Yang, and N.~Li, ``Is domain highlighting actually
  helpful in identifying phishing web pages?,'' {\em Hum. Factors}, vol.~59,
  no.~4, pp.~640--660, 2017.

\bibitem{pfeffel2019user}
K.~Pfeffel, P.~Ulsamer, and N.~M{\"u}ller, ``Where the user does look when
  reading phishing mails--an eye-tracking study,'' in {\em Int. Conf. on
  Human-Computer Interaction}, pp.~277--287, Springer, 2019.

\bibitem{canfield2016quantifying}
C.~I. Canfield, B.~Fischhoff, and A.~Davis, ``Quantifying phishing
  susceptibility for detection and behavior decisions,'' {\em Human factors},
  vol.~58, no.~8, pp.~1158--1172, 2016.

\bibitem{canfield2018setting}
C.~I. Canfield and B.~Fischhoff, ``Setting priorities in behavioral
  interventions: An application to reducing phishing risk,'' {\em Risk
  Analysis}, vol.~38, no.~4, pp.~826--838, 2018.

\bibitem{HUANG2022}
Y.~Huang, L.~Huang, and Q.~Zhu, ``Reinforcement learning for feedback-enabled
  cyber resilience,'' {\em Annual Reviews in Control}, 2022.

\bibitem{shi2019exploring}
B.~Shi, G.~Liu, H.~Qiu, Z.~Wang, Y.~Ren, and D.~Chen, ``Exploring voluntary
  vaccination with bounded rationality through reinforcement learning,'' {\em
  Physica A: Statistical Mechanics and its Applications}, vol.~515,
  pp.~171--182, 2019.

\bibitem{sanjab2020game}
A.~Sanjab, W.~Saad, and T.~Ba{\c{s}}ar, ``A game of drones: Cyber-physical
  security of time-critical uav applications with cumulative prospect theory
  perceptions and valuations,'' {\em IEEE Transactions on Communications},
  vol.~68, no.~11, pp.~6990--7006, 2020.

\bibitem{RN661}
L.~Huang and Q.~Zhu, ``Combating informational denial-of-service ({IDoS})
  attacks: Modeling and mitigation of attentional human vulnerability,'' in
  {\em International Conference on Decision and Game Theory for Security},
  pp.~314--333, Springer, 2021.

\bibitem{huang2021radams}
L.~Huang and Q.~Zhu, ``Radams: Resilient and adaptive alert and attention
  management strategy against informational denial-of-service ({IDoS})
  attacks,'' {\em arXiv preprint arXiv:2111.03463}, 2021.

\bibitem{lin2011does}
E.~Lin, S.~Greenberg, E.~Trotter, D.~Ma, and J.~Aycock, ``Does domain
  highlighting help people identify phishing sites?,'' in {\em Proceedings of
  the SIGCHI Conference on Human Factors in Computing Systems}, pp.~2075--2084,
  2011.

\bibitem{akhawe2013alice}
D.~Akhawe and A.~P. Felt, ``Alice in warningland: A large-scale field study of
  browser security warning effectiveness,'' in {\em 22nd {USENIX} Security
  Symposium ({USENIX} Security 13)}, pp.~257--272, 2013.

\bibitem{sheng2010falls}
S.~Sheng, M.~Holbrook, P.~Kumaraguru, L.~F. Cranor, and J.~Downs, ``Who falls
  for phish? a demographic analysis of phishing susceptibility and
  effectiveness of interventions,'' in {\em Proceedings of the SIGCHI
  conference on human factors in computing systems}, pp.~373--382, 2010.

\bibitem{liebling2014privacy}
D.~J. Liebling and S.~Preibusch, ``Privacy considerations for a pervasive eye
  tracking world,'' in {\em Proceedings of the 2014 ACM International Joint
  Conference on Pervasive and Ubiquitous Computing: Adjunct Publication},
  pp.~1169--1177, 2014.

\bibitem{kroger2019does}
J.~L. Kr{\"o}ger, O.~H.-M. Lutz, and F.~M{\"u}ller, ``What does your gaze
  reveal about you? on the privacy implications of eye tracking,'' in {\em IFIP
  International Summer School on Privacy and Identity Management},
  pp.~226--241, Springer, 2019.

\bibitem{posner2016attention}
M.~I. Posner and O.~S. Marin, {\em Attention and performance XI}.
\newblock Routledge, 2016.

\bibitem{nasser2020role}
G.~Nasser, B.~W. Morrison, P.~Bayl-Smith, R.~Taib, M.~Gayed, and M.~W. Wiggins,
  ``The role of cue utilization and cognitive load in the recognition of
  phishing emails,'' {\em Frontiers in big Data}, p.~33, 2020.

\bibitem{ackerley2022errors}
M.~Ackerley, B.~Morrison, K.~Ingrey, M.~Wiggins, P.~Bayl-Smith, N.~Morrison,
  {\em et~al.}, ``Errors, irregularities, and misdirection: Cue utilisation and
  cognitive reflection in the diagnosis of phishing emails,'' {\em Australas.
  J. Inf. Syst.}, vol.~26, 2022.

\bibitem{frazier2018bayesian}
P.~I. Frazier, ``Bayesian optimization,'' in {\em Recent Advances in
  Optimization and Modeling of Contemporary Problems}, pp.~255--278, INFORMS,
  2018.

\bibitem{janisse1973pupil}
M.~P. Janisse, ``Pupil size and affect: A critical review of the literature
  since 1960.,'' {\em Canadian Psychologist/Psychologie canadienne}, vol.~14,
  no.~4, p.~311, 1973.

\bibitem{kang2014pupil}
O.~E. Kang, K.~E. Huffer, and T.~P. Wheatley, ``Pupil dilation dynamics track
  attention to high-level information,'' {\em PloS one}, vol.~9, no.~8,
  p.~e102463, 2014.

\bibitem{van1987simulated}
P.~J. Van~Laarhoven and E.~H. Aarts, ``Simulated annealing,'' in {\em Simulated
  annealing: Theory and applications}, pp.~7--15, Springer, 1987.

\bibitem{holland2013complex}
C.~D. Holland and O.~V. Komogortsev, ``Complex eye movement pattern biometrics:
  the effects of environment and stimulus,'' {\em IEEE Transactions on
  Information Forensics and Security}, vol.~8, no.~12, pp.~2115--2126, 2013.

\end{thebibliography}

\begin{IEEEbiography}[{\includegraphics[width=1in,height=1.25in,clip,keepaspectratio]{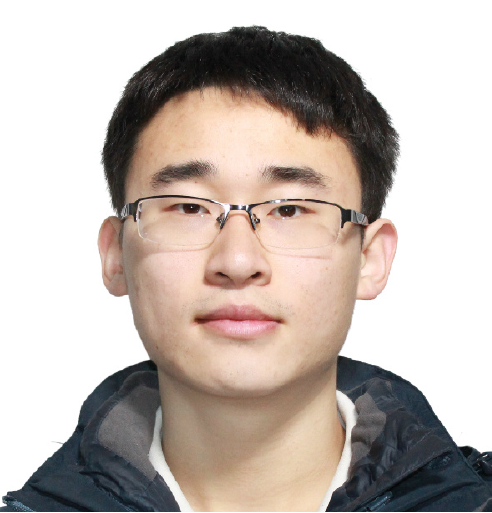}}]{Linan Huang} 
received the B.Eng. degree (Hons.) in Electrical Engineering from Beijing Institute of Technology, China, in 2016 and the Ph.D. degree in electrical engineering from New York University (NYU), Brooklyn, NY, USA, in 2022. 
%He is currently pursuing a Ph.D. degree at the Laboratory for Agile and Resilient Complex Systems, Tandon School of Engineering, New York University, NY, USA.
His research interests include dynamic decision-making of the multi-agent system, mechanism design, artificial intelligence, security, and resilience for cyber-physical systems. 
\end{IEEEbiography}
\begin{IEEEbiography}[{\includegraphics[width=1in,height=1.25in,clip,keepaspectratio]{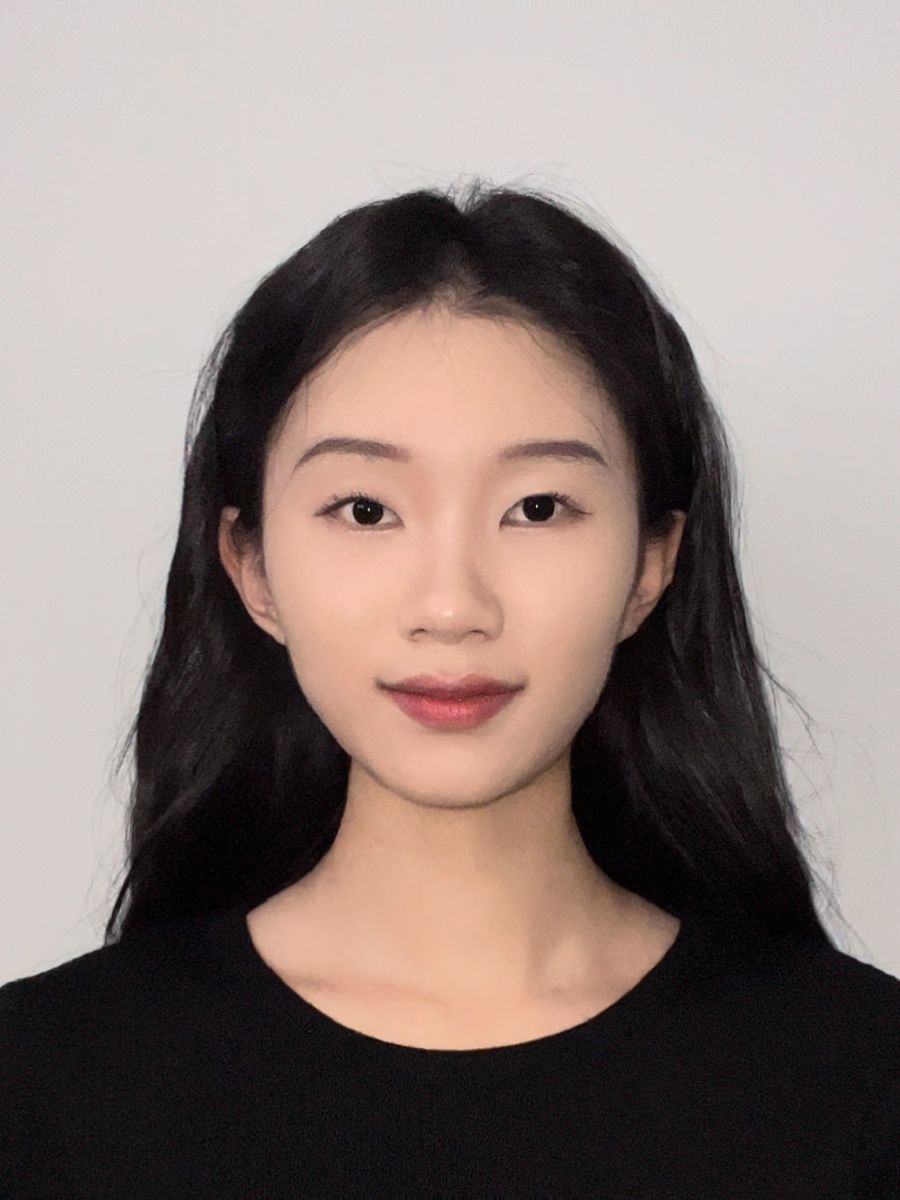}}]{Shumeng Jia} received the B.Eng. degree in rail traffic signaling and control from Bejing Jiaotong University, China, in 2020, and the M.S. degree in electrical engineering from New York University in 2022. She worked at the Laboratory for Agile and Resilient Complex Systems, Tandon School of Engineering, New York University, NY, USA, as a graduate assistant while working on her graduate degree. Her past research has focused on machine learning and epileptiform EEG data.
\end{IEEEbiography}
\begin{IEEEbiography}[{\includegraphics[width=1in,height=1.25in,clip,keepaspectratio]{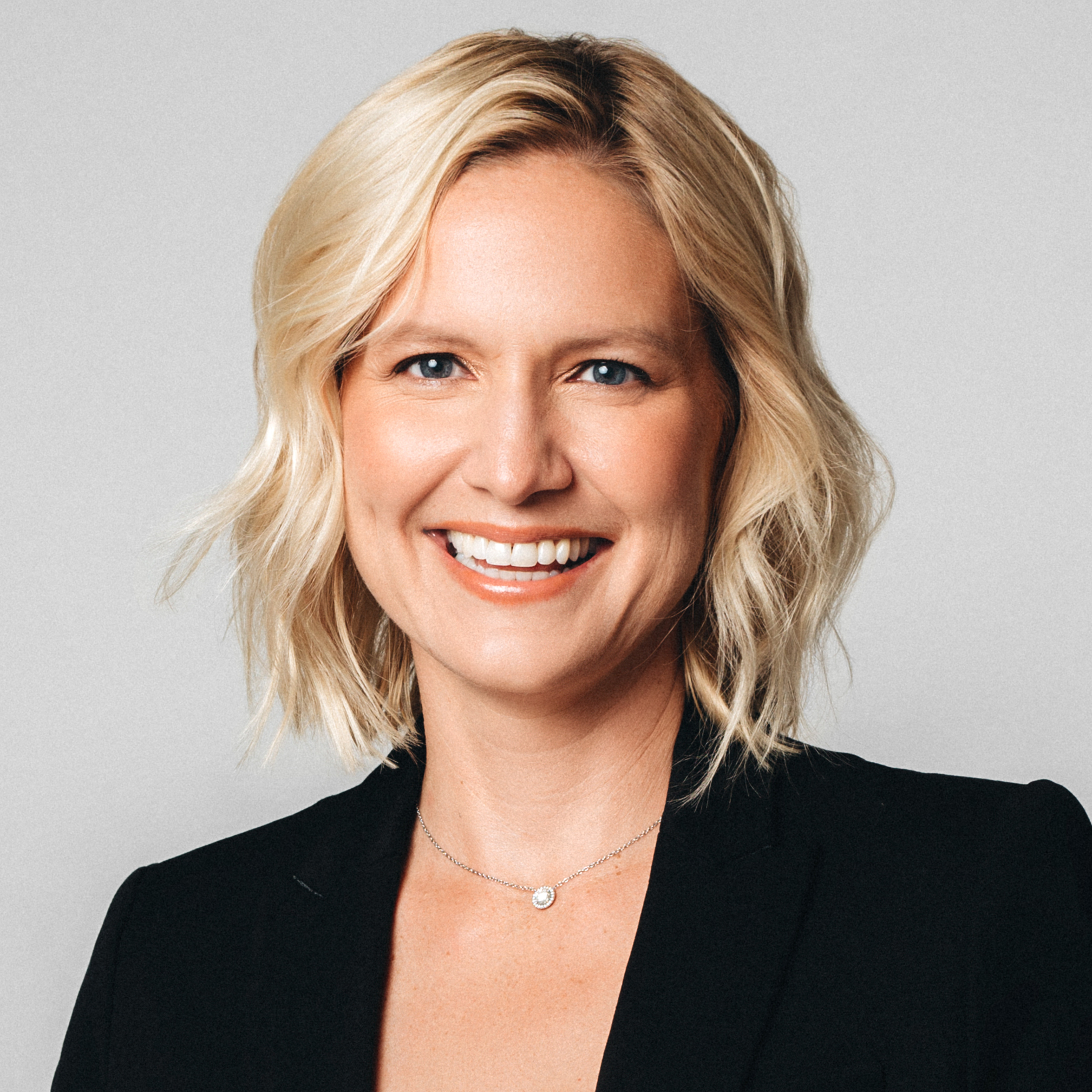}}]{Emily Balcetis}
received a BA (honors) in Psychology and a BFA in Music Performance from the University of Nebraska at Kearney in 2001 and a PhD in Social Psychology from Cornell University in 2006. She is currently an Associate Professor of Psychology at New York University (NYU), and faculty affiliate of NYU’s Institute for Human Development and Social Change. Her current research interests include motivation, decision-making, and visual experience. 
\end{IEEEbiography}
\begin{IEEEbiography}[{\includegraphics[width=1in,height=1.25in,clip,keepaspectratio]{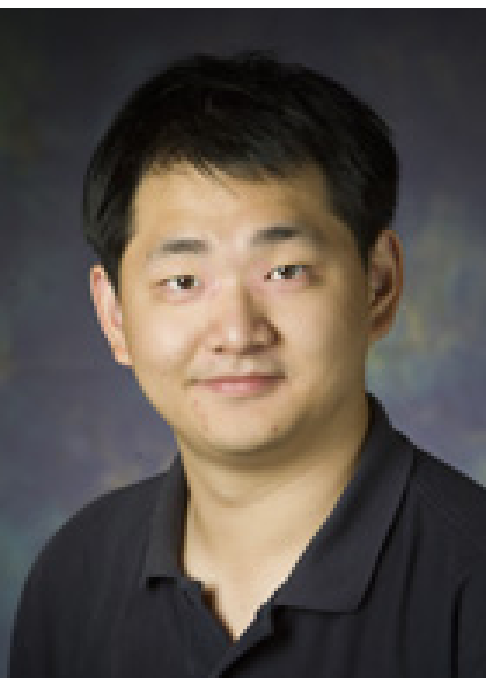}}]{Quanyan Zhu}
(SM’02-M’14) %IEEE student member 2002 and member 2014
received B. Eng. in Honors Electrical Engineering from McGill University in 2006, M. A. Sc. from the University of Toronto in 2008, and Ph.D. from the University of Illinois at Urbana-Champaign (UIUC) in 2013. 
After stints at Princeton University, he is currently an associate professor at the Department of Electrical and Computer Engineering, New York University (NYU). He is an affiliated faculty member of the Center for Urban Science and Progress (CUSP) and Center for Cyber Security (CCS) at NYU. His current research interests include game theory, machine learning, cyber deception, and cyber-physical systems.
\end{IEEEbiography}

\vfill

\end{document}